\documentclass[preprint2]{aastex7}

\usepackage{natbib}
\usepackage{amssymb,amsmath}
\newcommand*{\com}[1]{\textcolor{magenta}{#1}}

\newcommand{\rmu}{rad m$^{-2}$} 

\begin{document}

\title{RM-Tools: Software for Analyzing Polarized Radio Spectra}

\correspondingauthor{Cameron~L.~Van~Eck}

\author[0000-0002-7641-9946]{Cameron~L.~Van~Eck}\altaffiliation{The first two authors contributed equally to this work.}
\affiliation{Research School of Astronomy \& Astrophysics, The Australian National University, Canberra, ACT 2611, Australia}
\affiliation{Dunlap Institute for Astronomy and Astrophysics, University of Toronto, 50 St.\ George Street, Toronto, ON M5S 3H4, Canada}
\email[show]{cameron.vaneck@anu.edu.au}

\author{Cormac~R.\,Purcell}\altaffiliation{The first two authors contributed equally to this work.}
\affiliation{School of Computer Science and Engineering, UNSW Sydney, Sydney, NSW 2052, Australia}
\affiliation{Sydney Institute for Astronomy (SIfA), School of Physics, The University of Sydney, NSW 2006, Australia}
\email{cormac.r.purcell@gmail.com}


\author[0000-0003-0520-0696]{Lerato~Baidoo}
\affiliation{Dunlap Institute for Astronomy and Astrophysics, University of Toronto, 50 St.\ George Street, Toronto, ON M5S 3H4, Canada}
\email{mll.sebokolodi@gmail.com}

\author[0000-0001-9472-041X]{Alec~J.~M.~Thomson}
\affiliation{SKA Observatory, SKA-Low Science Operations Centre, 26 Dick Perry Avenue, Kensington WA 6151, Australia}
\affiliation{ATNF, CSIRO Space \& Astronomy, PO Box 1130, Bentley, WA 6102, Australia}
\email{alec.thomson@skao.int}

\author[0000-0003-0742-2006]{Yik~Ki~Ma}
\affiliation{Max-Planck-Institut f\"ur Radioastronomie, Auf dem H\"ugel 69, 53121 Bonn, Germany}
\affiliation{Research School of Astronomy \& Astrophysics, The Australian National University, Canberra, ACT 2611, Australia}
\email{ykma@mpifr-bonn.mpg.de}

\author{Lindsey~Oberhelman}
\affiliation{Research School of Astronomy \& Astrophysics, The Australian National University, Canberra, ACT 2611, Australia}
\email{Lindsey.Oberhelman@anu.edu.au}

\author[0000-0002-5815-8965]{Erik~Osinga}
\affiliation{Dunlap Institute for Astronomy and Astrophysics, University of Toronto, 50 St.\ George Street, Toronto, ON M5S 3H4, Canada}
\email{erik.osinga@utoronto.ca}

\author[0009-0004-7773-1618]{Shannon~Vanderwoude}
\affiliation{Dunlap Institute for Astronomy and Astrophysics, University of Toronto, 50 St.\ George Street, Toronto, ON M5S 3H4, Canada}
\email{shannon.vanderwoude@mail.utoronto.ca}

\author[0000-0001-7722-8458]{Jennifer~L.~West}
\affiliation{Dominion Radio Astrophysical Observatory, Herzberg Research Centre for Astronomy and Astrophysics, National Research Council Canada, PO Box 248, Penticton, BC V2A 6J9, Canada}
\affiliation{Dunlap Institute for Astronomy and Astrophysics, University of Toronto, 50 St.\ George Street, Toronto, ON M5S 3H4, Canada}
\email{jennifer.west@nrc-cnrc.gc.ca}

\author{Shinsuke~Ideguchi}
\affiliation{National Astronomical Observatory of Japan, 2-21-1 Osawa, Mitaka, Tokyo 181-8588, Japan}
\email{shinsuke.ideguchi@nao.ac.jp}

\author[0000-0002-5811-0136]{Dylan M. Par\'e}
\affiliation{Department of Physics, Villanova University, 800 E. Lancaster Ave., Villanova, PA 19085, USA}
\email{dylanpare@gmail.com}

\author{Jane~F.~Kaczmarek}
\affiliation{Sydney Institute for Astronomy (SIfA), School of Physics, The University of Sydney, NSW 2006, Australia}
\affiliation{CSIRO Parkes Observatory, CSIRO Space \& Astronomy, PO Box 276, Parkes, NSW 2870, Australia}
\email{jane.kaczmarek@skao.int}

\author{Tony~Willis}
\affiliation{Dominion Radio Astrophysical Observatory, Herzberg Research Centre for Astronomy and Astrophysics, National Research Council Canada, PO Box 248, Penticton, BC V2A 6J9, Canada}
\email{tony.willis.research@gmail.com}

\author[0000-0001-9399-5331]{Takuya~Akahori}
\affiliation{Mizusawa VLBI Observatory, National Astronomical Observatory Japan, 2-21-1 Osawa, Mitaka, Tokyo 181-8588, Japan}
\affiliation{Sydney Institute for Astronomy (SIfA), School of Physics, The University of Sydney, NSW 2006, Australia}
\email{takuya.akahori@nao.ac.jp}

\author{Craig~S.~Anderson}
\affiliation{Research School of Astronomy \& Astrophysics, The Australian National University, Canberra, ACT 2611, Australia}
\email{craig.anderson@anu.edu.au}

\author[0000-0002-3382-9558]{B.~M.~Gaensler}
\affiliation{Division of Physical and Biological Sciences, University of California, Santa Cruz, 1156 High Street, Santa Cruz, CA 95064, USA}
\affiliation{Dunlap Institute for Astronomy and Astrophysics, University of Toronto, 50 St.\ George Street, Toronto, ON M5S 3H4, Canada}
\affiliation{David A. Dunlap Department of Astronomy and Astrophysics, University of Toronto, 50 St. George Street, Toronto, ON M5S 3H4, Canada}
\affiliation{Sydney Institute for Astronomy (SIfA), School of Physics, The University of Sydney, NSW 2006, Australia}
\email{gaensler@ucsc.edu}

\author[0000-0002-3968-3051]{Shane~O'Sullivan}
\affiliation{Departamento de Física de la Tierra y Astrofísica \& IPARCOS-UCM, Universidad Complutense de Madrid, 28040 Madrid, Spain}
\affiliation{Sydney Institute for Astronomy (SIfA), School of Physics, The University of Sydney, NSW 2006, Australia}
\email{s.p.osullivan@ucm.es}

\author[0000-0002-3464-5128]{Xiaohui~Sun}
\affiliation{School of Physics and Astronomy, Yunnan University, Kunming 650091, People’s Republic of China}
\affiliation{Sydney Institute for Astronomy (SIfA), School of Physics, The University of Sydney, NSW 2006, Australia}
\email{xhsun@ynu.edu.cn}


\author[0000-0002-7098-7512]{Ariel~D.~Amaral}
\affiliation{David A. Dunlap Department of Astronomy and Astrophysics, University of Toronto, 50 St. George Street, Toronto, ON M5S 3H4, Canada}
\affiliation{Dunlap Institute for Astronomy and Astrophysics, University of Toronto, 50 St.\ George Street, Toronto, ON M5S 3H4, Canada}
\email{ariel.amaral@mail.utoronto.ca}

\author[0000-0002-3369-1085]{C.~J.~Riseley}
\affiliation{Astronomisches Institut der Ruhr-Universit\"{a}t Bochum (AIRUB), Universit\"{a}tsstra{\ss}e 150, 44801 Bochum, Germany}
\email{riseley@astro.ruhr-uni-bochum.de}

\author[0000-0003-2623-2064]{Jeroen~Stil}
\affiliation{Department of Physics and Astronomy, the University of Calgary, 2500 University Drive NW, Calgary, AB T2N 1N4, Canada}
\email{jstil@ucalgary.ca}

\author[0000-0002-2218-5638]{Xiang~Zhang}
\affiliation{LIRA, Observatoire de Paris, Université PSL, CNRS, Sorbonne Université, Université Paris Cité, 5 place Jules Janssen, 92195 Meudon, France}
\email{Xiang.Zhang@obspm.fr}

\begin{abstract}
Polarization observations using modern radio telescopes cover large numbers of frequency channels over broad bandwidths, and require advanced techniques to extract reliable scientific results.  We present {\em RM-Tools}, analysis software for deriving polarization properties, such as Faraday rotation measures, from spectropolarimetric observations of linearly polarized radio sources. The software makes use of techniques such as rotation measure synthesis and QU-model fitting, along with many features to simplify and enhance the analysis of radio polarization data. RM-Tools is currently the main software that large-area polarization sky surveys such as POSSUM and VLASS deploy for science-ready data processing. The software code is freely available online and can be used with data from a wide range of telescopes.
\end{abstract}


\section{Introduction}\label{sec:intro}

Magnetic fields are known to play a significant role in many astrophysical environments, from planets and stars \citep[e.g.,][]{McIntyre2019, Brun2017} to the interstellar medium \citep[e.g.,][]{Seta2025} to galaxies \cite[e.g.,][]{Irwin2024}, galaxy groups \citep[e.g.,][]{Anderson2024}, and galaxy clusters \citep[e.g.][]{Osinga2025}. Understanding the magnetic field properties in these environments is crucial to understanding the behaviour and evolution of these systems. This requires the acquisition of high-quality data and the development of robust analysis methods to provide strong constraints for models and simulations.

Almost all methods used to measure the strength and direction of magnetic fields in astrophysical contexts rely on detecting polarized emission. There are two physical processes that relate magnetic fields to radio-band continuum emission: synchrotron radiation, which produces polarized emission, and Faraday rotation, which modulates the polarized signal of propagating emission. Polarized radio emission can be analyzed to determine properties of the magnetic field of synchrotron-emitting and Faraday-rotating astrophysical environments.

Analysis of Faraday rotation in radio observations has traditionally, up until approximately ten years ago, been performed using bespoke software tools written by individual astronomers for their own needs \citep[e.g.,][]{Oren1995,Brown2003,vanEck2011}. More recently the trend towards increasingly sophisticated and computationally complex polarization analysis methods and the increasing awareness of open source software ideas in the astronomical community has led to the development of several publicly released software tools for radio polarization analysis. Examples of this include Michiel Brentjens' {\em rm-synthesis}\footnote{https://github.com/brentjens/rm-synthesis}, Michael Bell's {\em pyrmsynth}\footnote{https://github.com/mrbell/pyrmsynth}, and {\em cuFFS}\footnote{https://github.com/sarrvesh/cuFFS} \citep{Sridhar2018}. These tools provide significant value to the community, but in a relatively narrow scope of capabilities since each was developed by a single researcher for a comparatively short period of time, often with a specific use case in mind and without extensive documentation.

During the early development of data analysis pipelines for the Polarization Sky Survey of the Universe's Magnetism \citep[POSSUM;][]{Gaensler2025} and Very Large Array Sky Survey \citep[VLASS;][]{Lacy2020}, we recognized that the tools being produced could provide significant value to the astronomical community by extending beyond the capabilities of the existing radio polarization analysis tools. This resulted in the development of the software package described here, {\em RM-Tools}: a radio polarization analysis toolkit that offers a number of features beyond the scope of existing packages.

In Section~\ref{sec:pol} we review the theoretical basis for polarimetric analysis, specifically Faraday rotation measurement. In Section~\ref{sec:algorithms} we present the algorithms used in the RM-Tools package, including a new metric  for measuring Faraday complexity. Section~\ref{sec:interface} describes the top-level tools and interfaces with which users can interact with the package. Section~\ref{sec:tests} shows tests of some of the features within RM-Tools. Finally, in Section~\ref{sec:summary} we summarize the features of RM-Tools and plans for its future.


\section{Polarization and Faraday rotation principles}\label{sec:pol}
In this section we review the aspects of the mathematical and theoretical basis of radio polarization that are necessary for understanding RM-Tools' major features.

\subsection{Basic polarization parameters}
The polarization state of astrophysical radio emission is typically described using the Stokes parameters $I$, $Q$, $U$, and $V$.
From these, a number of further quantities can be derived: the complex polarization\footnote{Throughout this work, we will denote complex quantities with a tilde, the magnitude of that quantity using the same symbol without a tilde, and vector/array quantities with bold-face.}
$\tilde{\mathcal P} = Q + i U = {\mathcal P} e^{2i \psi}$ 
where $\mathcal P$ is the (linear) polarized intensity, defined as the modulus of the complex polarization, and $\psi$ is the polarization angle (also called the electric vector position angle; EVPA), defined as half the argument of the complex polarization. The (linear) polarized fraction is defined as $p = {\mathcal P} / I$.

\subsection{Faraday rotation}\label{sec:far_rot}
A detailed description of Faraday rotation can be found in the careful derivation by \citet{Ferriere2021}. The key quantity of interest is the Faraday depth, $\phi(l)$, which gives the strength of the Faraday rotation between a distance $l$ and the observer. 
The case of a line of sight consisting of a single source of polarized emission at a single distance and Faraday rotation caused by the medium in the foreground of the source is called {\em Faraday-simple} or Faraday-thin; the Faraday depth in this case is often called the rotation measure (RM). 

In general, lines of sight can be more complicated, with multiple sources of polarized emission at different distances (and thus different Faraday depths) or emission that is distributed over a range of distance and correspondingly over a range of Faraday depth; such cases are called Faraday-thick or Faraday-complex. Faraday-simple lines of sight can be analyzed fairly straightforwardly, e.g. with a linear fit to $\psi$ versus $\lambda^2$ \citep[e.g.,][]{Rand1994, Brown2003}, but complex lines of sight require more sophisticated analysis methods such as RM-synthesis or QU-fitting.
\subsection{RM-synthesis}
RM-synthesis transforms observed complex polarized spectra from the $\lambda^2$ domain to the Faraday depth, $\phi$, domain via a Fourier transform. The resulting (complex) function is called the Faraday dispersion function (FDF), which describes the distribution of polarized emission in the space of possible Faraday depths \citep{Burn1966,Brentjens2005}.

In practice, several modifications from the simple Fourier transform are required. First, the polarization $\tilde{\mathcal P}(\lambda^2)$ is measured for a series of discrete channels, $\tilde{\mathcal P}_k$ for channel $k \in (1..N_c)$ for an observation with $N_c$ channels, which results in the Fourier transform being converted to a discrete Fourier transform. Similarly, the Faraday depths are sampled at discrete values $\phi_j$. 

The discrete FDF is calculated via the sum
\begin{equation}\label{eq:fdf} { \tilde{\mathcal F}(\phi_j) =
  H\,\sum_{k=1}^{N_c}\,w_k\,\tilde{\mathcal
    P_k}\,e^{-2\,i\,\phi_j\,(\lambda^2_k-\lambda^2_0)}},
\end{equation}
where $w_i$ is a weighting function and $\lambda_0^2$ is the fiducial reference wavelength-squared, which is typically taken to be the weighted mean of the observed $\lambda^2$ channels.
Here ${ H=(\sum_{k=1}^{N_c}\,w_k)^{-1}}$ is a normalisation constant.  

The resulting ``dirty'' FDF is convolved with the sampling response ${\mathcal R}(\phi_{ j})$, called the rotation measure spread function (RMSF) or rotation measure transfer function (RMTF). This may be deconvolved, for example, using the \cite{Hogbom1974} {\scriptsize CLEAN} algorithm RM-clean (see \S\ref{sec:rmclean} and \citealt{Heald2009}).

The complex RMSF may be calculated from the $\lambda^2$ sampling of the observations and the weight applied to each channel via
\begin{equation}\label{eq:rmsf}
 {  \tilde{\mathcal R}(\phi_j) = H\,\sum_{k=1}^N\,w_k\,e^{-2\,i\,\phi_j\,(\lambda^2_k-\lambda^2_0)}}.
\end{equation}
The full width at half maximum intensity (FWHM) of the RMSF is the most common value reported to characterize the RMSF.

RM-synthesis is often performed on the fractional polarization spectra $q(\lambda^2)=Q(\lambda^2)/I_{\rm mod}(\lambda^2)$ and $u(\lambda^2)=U(\lambda^2)/I_{\rm mod}(\lambda^2)$, which have been normalized by a model Stokes\,$I$ spectrum to remove the effect of spectral slope and curvature. If this normalization is not performed, strong spectral variations (e.g., steep power-law spectra) can introduce artifacts in the FDF at the few-percent level \citep{Brentjens2005}. The total intensity model spectrum is typically obtained by fitting a power-law or polynomial to the total intensity data; direct normalization with measured channel values is typically not used as the limited total intensity signal-to-noise ratio in individual channels significantly increases the noise in the fractional polarization spectra. When this normalization is performed, the resulting FDF is also in units of fractional polarization; it can be multiplied by a fiducial Stokes $I$ value (e.g., the model Stokes $I$ at a chosen reference frequency/wavelength) to convert the FDF back to the same units as the input Stokes parameters.

The theoretical noise value, $\sigma(\tilde{\mathcal F})_{\rm th}$,\footnote{Throughout this work we will use $\sigma(X)$ to denote an estimated measurement uncertainty in quantity $X$, while $\sigma$ without brackets will be used for other quantities, such as the widths of distributions (e.g., channel noise).}
in the FDF, can be calculated via
\begin{equation}
  \sigma({\tilde{\mathcal F}})_{\rm th}^2=H^2 \sum_{k=1}^N  (w_k \sigma_{{\rm qu},k})^2,
\label{eq:sigma_fth}
\end{equation}
where $\sigma_{{\rm qu},k}$ is the estimated noise of the $k^{\rm th}$ frequency channel, computed by averaging the supplied noise in $Q$ and $U$ and dividing by the model Stokes $I$ value if this has also been applied to the FDF. This value represents the expected 1-$\sigma$ variation in Stokes $Q$ and $U$ (real and imaginary components) of the FDF. If a data set has substantively different noise properties between $Q$ and $U$, this behavior is not captured by this method and such differences are not propagated into later stages of the uncertainty analysis.

RM-synthesis has the advantage of not being affected by the n-$\pi$ ambiguity present when fitting discrete values of $\psi$ versus $\lambda^2$.\footnote{More specifically, n-$\pi$ ambiguity becomes high-amplitude sidelobes in the RMSF.} However, Faraday-thick emission can manifest as a resolved peak in $|{\mathcal F}(\phi)|$ (assuming sufficient $\lambda^2$ coverage) and is not easy to characterize or interpret. Depending on the wavelength-squared coverage of the input data, broad Faraday-depth features may be reduced in amplitude or entirely lost (strongly analogous to how radio interferometers have the ``missing short spacings'' which resolves out large-scale emission). This effect has been quantified in different ways \citep{Brentjens2005, Rudnick2023}, but depends on the smallest $\lambda^2$ value (highest frequency) sampled.
RM-synthesis also has the advantage of being non-parametric and does not contain any assumptions about the shape of the FDF; however meaningful physical properties of complex polarized sources are often more easily extracted by modeling the line-of-sight distributions of polarized sources and RM rather than performing RM-synthesis \citep[e.g.][]{Sun2015}.

\subsection{QU-fitting: modeling polarized spectra}
The term QU-fitting refers to any technique used to fit parametric models directly to the observed spectral data. Recently, this approach has been applied successfully to studies of various systems, including 
supernova remnants \citep{Shanahan2023}, 
diffuse Galactic emission \citep{Thomson2021, Mohammed2024}, 
and active galactic nuclei \citep{O'Sullivan2012,Anderson2016}
The general approach is to test a suite of parametric models against each dataset in a two-step procedure: 1) for each model, determine the parameters that best fit the data based on the $\chi^2$ test; 2) compare between best-fit models to select a preferred model by using an information criterion that penalizes over-complex models.

Models manifest as complex polarization vectors $\tilde{\mathcal P}(\lambda^2)$, as described in $\S$\ref{sec:pol_models}.
The fitting procedure can, in principle, be any nonlinear model fitting algorithm, but it should be noted that gradient descent-type algorithms tend not to perform well due to poor behavior of the $\chi^2$ surface far away from the correct parameters. Markov chain Monte Carlo (MCMC) and nested sampling methods have been used successfully \citep{Miyashita2019, Mckinven2021}, and have advantages in terms of providing additional information in the form of sampling the likelihood or posterior distributions of the model parameters. 

Model selection has been done using a variety of different metrics, including direct $\chi^2$ comparisons \citep{Kaczmarek2018}, the Akaike or Bayesian information criteria (AIC or BIC) \citep{O'Sullivan2017,Schnitzeler2018}, and Bayesian evidence \citep{Thomson2021}. While the $\chi^2$ is a direct measurement of the goodness-of-fit, the AIC and BIC are modifications of the $\chi^2$ that apply penalties for number of free parameters in the model (and thus disfavor complex models that do not significantly improve the fit). The Bayesian evidence, or marginal likelihood, provides a more sophisticated statistical tool for comparing multiple models.

Many different QU models have been proposed based on different physical conditions, such as different combinations of emitting and Faraday rotating regions, and different magnetic field configurations such as uniform or turbulent \citep[e.g.,][]{Burn1966,Tribble1991,Sokoloff1998}. In Appendix~\ref{sec:pol_models} we briefly describe the models that have been included in RM-Tools.

\subsection{Comparative merits of RM-synthesis and QU-fitting}
While both RM-synthesis and QU-fitting can be applied to the same data (polarized radio spectra), these methods have some differing properties and use cases. RM-synthesis is non-parametric and model-independent, and is also comparatively much less computationally expensive; these properties make it suitable for detection experiments, initial exploration of polarization properties, or for analysis of large datasets. RM-synthesis does not (currently) have well-defined methods for detailed analysis of Faraday-complex lines of sight. The combination of the RMSF, which acts as a point-spread function to ``blur out'' details in the FDF, and the missing large-scales, which suppress broad features, makes it difficult to confidently interpret a complex FDF in terms of physical structures.

Conversely, QU-fitting can directly extract model parameters that can have physical interpretations, provided that the correct model is selected. QU-fitting can often be more effective in identifying and characterizing Faraday complexity, especially at lower signal-to-noise levels (Oberhelman et al., submitted for publication). This comes with challenges in identifying if the correct model is selected (assuming that a given line of sight can even be modeled analytically; see e.g.\ \citealt{Basu2019}) and that the correct model parameters can be recovered; this may not happen reliably for all models and parameters (L. Oberhelman et al., submitted for publication).

Both methods share certain properties, mostly tied to the common input data they share. The ability to identify and characterize Faraday-complexity is strongly dependent on the wavelength-squared coverage of the data. When analyzing Faraday-simple sources, both methods achieve the same level of uncertainty in the derived polarization properties. For more complex sources, the measurement uncertainties are not always reliable from QU-fitting (L. Oberhelman et al., submitted for publication), while no corresponding analysis has yet been performed for RM-synthesis.


\section{RM-Tools - Algorithms}\label{sec:algorithms}

The RM-Tools software provides a toolkit for several different types of radio polarization analysis methods, including RM-synthesis, RM-{\scriptsize CLEAN}, and QU-fitting, along with additional algorithms for characterizing Faraday complexity, extracting polarized properties from FDFs, uncertainty estimation, and more. The software is implemented as a {\it Python 3} module that is available on PyPI\footnote{https://pypi.org/project/RM-Tools/}, GitHub\footnote{https://github.com/CIRADA-Tools/RM-Tools}, and Zenodo (doi:10.5281/zenodo.17851649) under an MIT licence.

In this section we describe the specific algorithms implemented in RM-Tools; the broader user-level interface for these algorithms will be described in \S\ref{sec:interface}.

\subsection{Stokes I modeling}
\label{sec:model_I}
As described in \S\ref{sec:far_rot}, it is common to normalize polarized spectra by the total intensity spectra to remove artifacts that can be introduced by spectral structure that is not related to the polarization behaviour (such as spectral index). RM-Tools does not require this step, but does offer the capability to perform this normalization. In RM-Tools this is achieved by dividing the Stokes~{\it Q} and~{\it U} spectra by model Stokes~{\it I} spectra (${I}_{\rm mod}(\nu)$), which have been constructed by fitting smooth functions to the data.
 Routines have been included to fit the supplied Stokes $I$ spectra and generate the Stokes {\it I} model used.

Two fitting functions have been implemented. The first is a (linear) polynomial of up to 5th order, of the form 
\begin{equation}
   I(\nu) = \sum_{n=0}^5 C_n \left(\frac{\nu}{\nu_0}\right)^n 
\end{equation}
with coefficients $C_n$ and reference frequency $\nu_0$. The second is a power-law function with up to four curvature terms, equivalent to a log-log space polynomial of up to 5th order, of the form
\begin{equation}
I(\nu) = C_0 \left(\frac{\nu}{\nu_0}\right)^{C_1 + C_2 \log_{10}(\nu/\nu_0) + C_3 (\log_{10}(\nu/\nu_0))^2 + ...}    
\end{equation}
where $C_0$ and $C_1$ can be interpreted respectively as the Stokes $I$ value and spectral index at the reference frequency. The reference frequency is initially chosen to be the unweighted mean of the frequencies of all channels with non-NaN Stokes {\it I} values; after $\lambda^2_0$ is determined, the model parameters are rescaled to the final polarization reference frequency corresponding to $\nu_0 \equiv c/\lambda_0$. The fitting routine uses a Levenberg-Marquardt non-linear least-squares fitting algorithm to find the best-fit parameters and corresponding parameter uncertainties.

The linear polynomial model was implemented first and was found to have problems in low signal-to-noise regimes, wherein the best-fit model could contain negative Stokes {\it I} values which in turn flips the sign of the fractional Stokes Q and U values. The log-polynomial model was implemented to prevent this, although the fitting algorithm imposes no constraints on the parameter values and the $C_0$ parameter can still take negative values; in this case, the entire Stokes {\it I} model becomes negative. The log-polynomial is both less likely to have problems and is generally a better model for the Stokes {\it I} spectrum, so it is the recommended model; the linear-polynomial model has been left in for users that prefer it. For both models, in cases where the model contains at least one negative value, a flag is used to signal the presence of such occurrences to the user.

The order of the polynomial can either be selected by the user (up to fifth order), or set dynamically by the fitting routine. If dynamic order selection is used, the routine starts by fitting a zeroth order model, then iteratively increases the model order, fits the new model, and compares the new model to the previous using the AIC; when the AIC no longer decreases, the fitter stops and uses the model with the lowest AIC value. The highest allowed order is set by the user, up to fifth order. This procedure prevents over-fitting by disfavoring more complex models that do not significantly improve the fit. 

Division by the resulting Stokes {\it I} model produces fractional spectra $q(\nu)=Q(\nu)/I_{\rm mod}(\nu)$ and $u(\nu)=U(\nu)/I_{\rm mod}(\nu)$ that have similar noise properties to the original data. The uncertainties in the fractional spectra are computed by applying the same normalization to the uncertainties in {\it Q} and~{\it U}: $\sigma_{q}(\nu) = \sigma_{Q}(\nu)/I_\mathrm{mod}(\nu)$ and $\sigma_{u}(\nu) = \sigma_{U}(\nu)/I_\mathrm{mod}(\nu)$. No additional terms to account for uncertainties in the Stokes {\it I} model are included, as they are expected to not contribute significantly in the vast majority of cases; polarization fractions are usually low (5-10\%), implying that sources with a good signal-to-noise ratio in polarization will have a high signal-to-noise ratio in total intensity.

\subsection{RM-Synthesis}\label{sec:rmsynth}
The core RM-synthesis module takes in polarized spectra and calculates the corresponding FDF for each spectrum. The primary inputs are the Stokes $q(\lambda^2)$ and $u(\lambda^2)$ arrays\footnote{The implicit convention is to consider the spectra interchangeably as functions of frequency or $\lambda^2$; using either the channel center frequencies $\nu$ or the corresponding $\lambda^2 = (c/\nu)^2$ values as appropriate for the equation in question. This is valid for channels of infinitesimal bandwidth, but introduces an error with increasing per-channel fractional bandwidth; this effect is discussed in briefly in \citet{Brentjens2005} and in detail by \citet{Schnitzeler2015}.}
(or $Q$ and $U$, if Stokes $I$ normalization has not been performed), a vector of the $\lambda^2$ values of all channels, and a vector of the Faraday depth values at which the FDF is to be calculated. Optionally, a vector of channel weights can be supplied (otherwise all channels are weighted uniformly) as well as the reference wavelength-squared value $\lambda^2_0$ (otherwise calculated from Eq.~32 of \citealt{Brentjens2005}).

The input spectra may be 1D, 2D, or 3D arrays, corresponding to single spectra, a list of spectra, or an image cube, but the $\lambda^2$, Faraday depth, and weight vectors must be 1D arrays; these parameters are assumed to be uniform for all spectra in a given execution.

The FDF is directly calculated as the (non-uniform) discrete Fourier transform \citep{Barnett2019} given in Eq.~\ref{eq:fdf} -- the spectra are not resampled onto a uniformly sampled $\lambda^2$ grid to enable a Fast Fourier Transform.  The resulting FDF has the same dimensionality as the input spectra, but with the frequency axis replaced with the Faraday depth axis.

While the core function allows the user to specify an arbitrary vector of Faraday depth values, the user interfaces that invoke this function all provide a standard method to define the Faraday depth sampling. The standard form of the sampling is uniformly-spaced and symmetric about zero \rmu; this sampling is fully specified by two parameters: the spacing between samples, $\Delta\phi$, and the largest Faraday depth sampled, $\phi_{\rm max}$. If the user specifies these values, the Faraday depth sampling is a vector from $-\phi_{\rm max}$ to $+\phi_{\rm max}$ (inclusive) in steps of $\Delta\phi$.\footnote{If the supplied or calculated $\phi_{\rm max}$ is not an integer multiple of $\Delta\phi$, it is rounded to the nearest multiple.}
If the user does not supply one or both of these parameters, they are calculated as follows. 

The theoretical FWHM of the RMSF is calculated using either the equation ${\rm FWHM} = 3.8 / (\lambda^2_{\rm max} - \lambda^2_{\rm min})$ \citep{Dickey2019} when $\lambda^2_0$ is defined as per \citet{Brentjens2005} or ${\rm FWHM} = 2 / (\lambda^2_{\rm max} + \lambda^2_{\rm min})$  \citep{Rudnick2023} when $\lambda^2_0$ is chosen to be zero, where $\lambda^2_{\rm max}$ and $\lambda^2_{\rm min}$ are the largest and smallest values respectively of the channel wavelength-squared. Both equations assume that the $\lambda^2$ range between $\lambda^2_{\rm min}$ and $\lambda^2_{\rm max}$ are fully and uniformly sampled; observations with frequency gaps will have more complicated RMSF shapes but are not generally expected to be narrower than the ideal uniform case, while observations with high fractional bandwidth and uniform sampling in frequency will also have a modified shape (due to nonuniform sampling in $\lambda^2$). The Faraday depth sampling step size is set as the FWHM divided by an oversampling factor that can be supplied by the user or kept at the default value of 10.

The largest Faraday depth sampled is, when not supplied by the user, set to be the larger of two possible values: 10 times the FWHM calculated above, or the 50\% bandwidth depolarization threshold. This threshold is calculated as defined by Eq.~15 of \citet{Brentjens2005} as $\sqrt{3}/\delta\lambda^2_{\rm min}$ where $\delta\lambda^2_{\rm min}$ is the width in wavelength-squared of the lowest frequency channel.

We note that the default values on both sampling parameters are very conservative: the default step size is quite small (much smaller than the resolution), and the default maximum Faraday depth is typically quite large (in some cases larger than the range of astrophysically expected). The resulting FDFs can contain unnecessarily large numbers of Faraday depth samples which may increase computational and data volume costs; users are encouraged to consider whether the default values best suit their data and goals.

If Stokes $I$ normalization was performed within RM-Tools, the computed FDF is multiplied by a fiducial Stokes $I$ value to convert the FDF back into the same units as the input data. This Stokes $I$ value is determined as the value of the Stokes $I$ model at the polarization reference frequency $\nu_0$. The resulting FDF has units of the spectra unit per RMSF (i.e., if the input spectra have units of Jy/beam, the FDF has units of Jy/beam/RMSF); quantities derived from the FDF amplitude (such as polarized intensity or noise amplitudes) have the same unit.

As RM-Tools relies on a non-uniform fast Fourier transform (nuFFT), it has very relaxed requirements for the sampling of the input spectra. There is no requirement for uniform sampling in either the $\lambda^2$ or frequency domains, nor is there a requirement for equal channel widths in either domain. The resulting FDF also has no strict requirements for the sampling step size or Faraday depth range; by convention it is uniformly sampled but this is not required. The penalty for such freedom is that FDFs computed by RM-Tools are not invertible -- it is not possible to directly transform an FDF back into the frequency domain and recover the original spectrum.

\subsection{RM-clean}
\label{sec:rmclean}
In many cases, one wishes to deconvolve the dirty FDF to try to remove the effect of the RMSF. The simplest and most popular deconvolution method for FDFs is to use a variation of the {\scriptsize CLEAN} algorithm developed by \citet{Hogbom1974} for use with aperture synthesis imaging. \citet{Heald2009} first demonstrated that the {\scriptsize CLEAN} algorithm could be applied to Faraday dispersions functions (``RM-clean'') and presented the full recipe in Appendix A of that paper. The RM-clean algorithm used in RM-Tools is broadly similar to that of \cite{Heald2009}, with a number of small changes. 

The RM-clean algorithm works by repeatedly subtracting a scaled version of $\mathcal{R}(\phi)$ shifted to the location and polarization angle (phase) of the peak sample in $|\mathcal{F}(\phi)|$. The subtraction is performed iteratively on the residual until all remaining samples are below some threshold, or until a user-defined maximum allowed number of iterations is reached. The user can specify the threshold in units of polarized intensity or as a multiple of the theoretical noise value. The scaling of the subtracted RMSF is set by the peak amplitude of the FDF and a gain parameter set by the user (typically 10\%).

RM-Tools implements the option of a two-stage clean, with separate thresholds. The initial stage searches the entire FDF for the strongest peak, and performs RM-clean as described above until reaching the first threshold. After this, RM-clean continues but only searches for peaks within a small window around previously identified clean components; the window width is equal to the FWHM of the RMSF. The windowed clean continues until the second threshold is met. The purpose of this windowed clean is to allow deeper cleaning of complex Faraday spectra while reducing the risk that clean components are placed on bright noise peaks or artifacts away from the expected polarized signal \citep{Thomson2023}; this is directly analogous to the windowed CLEAN used in interferometric imaging \citep{cotton2008evla116}. The thresholds for both stages of cleaning can be set by the user, either in the same units as the FDF or as a multiple of the theoretical noise level.

The amplitudes, phases, and positions of the detected peaks are accumulated on a `clean component' spectrum.
After the clean threshold is reached, the clean component spectrum is convolved with a Gaussian model of the main lobe of the RMSF and the result added to the last residual to form the `restored' or `clean' FDF. 

\subsection{Measuring properties of the FDF}
\label{sec:properties}
Once an FDF is computed, it remains necessary to analyze it to attempt to extract useful information about the line-of-sight polarized emission and Faraday rotation. There is no general prescription for this process as FDFs can be, in principle, arbitrarily complex. RM-Tools has a function that determines the properties of the highest amplitude peak in the FDF; in the case of a Faraday-simple FDF this should determine all the useful properties (e.g., RM, polarized intensity, and polarization angle) contained in the FDF, but as the FDF becomes increasingly Faraday complex many of these quantities become increasingly poorly defined and/or determined. Users are required to use their discretion in determining if the derived properties are meaningful and reliable.

RM-Tools begins by finding the FDF sample with the highest amplitude and recording its position in the FDF (i.e., taking the modulus of the complex FDF and finding the largest value). To determine the Faraday depth of the peak with more precision than the sampling step size, a 3-point parabolic fit is performed to the modulus of the FDF, taking the values of the peak and the points to either side of the peak. From this fit, the position (Faraday depth, $\phi_{\rm peak}$) and amplitude (polarized intensity, $P_{\rm peak}$) of the peak are extracted.

An estimate of the noise level in the FDF ($\sigma_{\rm FDF}$) is constructed by masking out the section of the FDF around the peak (equal to twice the FWHM of the RMSF on either side of the peak), computing the median absolute deviation from the median (MADFM) of a vector formed by concatenating the real and imaginary components of the FDF, and then normalizing the MADFM to an equivalent Gaussian $\sigma$ value.\footnote{Assuming that the underlying noise is predominantly Gaussian, relationship is $\mathrm{MAD} = \sigma\sqrt{2}\, \mathrm{erf}^{-1}(1/2) \approx0.6745\sigma$ \citep{Ruppert2010}.} 
The noise estimation is done on the real and imaginary components, rather than the modulus, to prevent problems with polarization bias and non-Gaussian statistics that are associated with the modulus. The purpose of the masking is to remove the majority of any real polarized signal from the noise estimation. Even with this mask, it is expected that for dirty FDFs the far sidelobes of any polarized signal present (or any polarized signal at Faraday depths far from the peak) will be present and may bias the noise estimate slightly high; this problem should not be present for cleaned FDFs.

The observed polarized intensity is corrected for polarization bias using the formula developed by \citet{George2012},
\begin{equation}\label{eqn:p_eff}
  {\mathcal P}_{\rm eff} = \sqrt{\mathcal{P}_{\rm peak}^2 - 2.3 \; \sigma({\mathcal F})_{\rm th}^2},
\end{equation}
but only for sources with a signal-to-noise ratio (S/N) greater than 5 (at low $S/N$ values, polarization bias becomes difficult to quantify, so we chose not to correct in such cases).

The Stokes $Q$ and $U$ values of the peak are derived by linearly interpolating the real and imaginary components respectively of the FDF to the location of the peak Faraday depth. The corresponding polarization angle ($\psi$) is defined in the usual way:\footnote{Computationally, this is defined using the \texttt{atan2} function, to ensure that the full range of possible polarization angles can be computed.}
\begin{equation}
  \psi_{\lambda_0} = \frac{1}{2}\,{\rm tan^{-1}}\,\left(\frac{{U|_{\phi_{\rm peak}}}}{Q|_{\phi_{\rm peak}}}\right)
\end{equation}
where the subscript $\lambda_0$ is used to emphasize that this is the ``in-band'' polarization angle at the polarization reference wavelength.

The in-band polarization angle and peak Faraday depth are used to estimate the {\it derotated} or {\it intrinsic} polarization angle:
\begin{equation}
  \psi_0 = \psi_{\lambda_0} - \phi_{\rm peak}\,\lambda_0^2.
\end{equation}
In the Faraday-simple case, this angle corresponds to polarization angle of the polarized emission before it was Faraday rotated. In the general Faraday-complex case, the derotated angle may not have any physical significance.

\subsection{Error analysis}
\label{sec:errors}
Uncertainty estimates for all of the measured FDF properties are also calculated. By default all the calculations are done using the theoretical FDF noise (Equation~(\ref{eq:sigma_fth})). The code calculates the empirically derived noise ($\sigma_{\rm FDF}$), and since all the uncertainty equations are linear in the noise they can be manually rescaled using the ratio of empirical noise to theoretical noise if desired. The 1D RM-clean tool outputs uncertainties calculated from both the theoretical noise and empirical noise (which have the additional suffix ``Observed'' in the output names).

The uncertainty in the fitted polarized intensity ${\mathcal P}_{\rm peak}$ is
\begin{equation}
  \sigma({\mathcal P}_{\rm peak}) = \sigma({\mathcal F})_{\rm th}
\end{equation}
where $\sigma({\mathcal F})_{\rm th}$ is defined in Equation~(\ref{eq:sigma_fth}). The bias-corrected polarized intensity is assumed to have the same error (no estimate for the uncertainty introduced by the bias correction is derived).

\citet{Brentjens2005} derived the uncertainty in the peak Faraday depth through analogy to linear regression, although this explicitly assumed equal noise and weights for all channels. We have found empirically (see \S\ref{sec:errortests}) that an analogy to fitting the centroid of a Gaussian and using the appropriate uncertainty equation from \citet{Condon1997} works very well and appears to be more flexible in accounting for channel weights by explicitly relating the uncertainty to the width of the RMSF. The resulting uncertainty in peak Faraday depth is:
\begin{equation}
    \sigma(\phi_{\rm peak}) = \frac{\sigma({\mathcal F})_{\rm th}}{{\mathcal P}_{\rm peak}}\, \frac{\rm FWHM}{2}
\end{equation}
where we note that the first fraction is the inverse of the signal-to-noise ratio in polarized intensity.

The uncertainty in the in-band polarization angle is given (in radians) by \citet{Brentjens2005} as:
\begin{equation}
  \sigma( \psi_{\lambda_0}) = \frac{1}{2}\,\frac{\sigma_{\rm
      QU}}{\mathcal{P}_{\rm peak}}.
\end{equation}
 and the uncertainty in the derotated polarization angle is given as:
\begin{equation}
  \sigma_{\psi_0} = \frac{\sigma({\mathcal F})_{\rm th}\, N_\nu}{4\,(N_{\nu}-2)\,{\mathcal P}_{\rm
      peak}^2}\,\left(\frac{N_{\nu}-1}{N_{\nu}} +
  \frac{\lambda_0^4}{\sigma_{\lambda^2}^2}\right),
\end{equation}
where $N_\nu$ is the number of frequency (or $\lambda^2$) channels and $\sigma_{\lambda^2}^2$ is the variance in the $\lambda^2$ distribution given by 
\begin{equation}\label{eqn:var_lambdaSq}
  \sigma_{\lambda^2}^2 = \frac{1}{N_{\nu}-1}\,{\textstyle\left[\sum_k\lambda_k^4 - N_{\nu}^{-1}\left(\sum_k\lambda_k^2\right)^2  \right]}.
\end{equation}

\subsection{Detecting Faraday complexity}
Faraday complexity can arise from a multitude of different physical scenarios (e.g., see $\S$\ref{sec:pol_models}, \citealt{Sokoloff1998}, and \citealt{Farnes2014}) and testing for all of these on all detected polarized sources is not generally practical. It is much simpler to ask the question: {\it Is the data consistent (or inconsistent) with a Faraday thin scenario?}. Two approaches to answering this question are used in the RM-Tools software. The first method operates directly on the Stokes~{\it q} and~{\it u} spectra, and tests whether the distribution of values deviates from that expected for a Faraday thin component. The second method tests if the Faraday depth dispersion of $|\tilde{\mathcal F}(\phi)|$ is significant compared to the width of the RMSF.

\subsubsection{Detecting complexity in a polarized spectrum - the $\sigma_{\rm add}$ metric}
\label{sec:complexity_qu}
We start by performing RM-synthesis on the fractional $q(\lambda^2)$ and~$u(\lambda^2)$ spectra\footnote{This method can be used without applying Stokes $I$ normalization, but any non-zero spectral index will introduce complexity in the model residuals, causing an increase in the value of $\sigma_{\rm add}$ even in a Faraday-simple source.}
to form the complex Faraday dispersion function $\tilde{\mathcal F}(\phi)$. We measure the peak Faraday depth $\phi_{\rm peak}$, intrinsic polarization angle $\psi_0$ and fractional polarization $p$ of the peak in $|\tilde{\mathcal F}(\phi)|$ and create a model Faraday thin spectrum using Equation~\ref{eqn:Faraday_model_thin}. The real and imaginary components of this model spectrum are subtracted from the data to form residual $q_{\rm res}(\lambda^2)$ and~$u_{\rm res}(\lambda^2)$ spectra. Any unexpected structure in the residual spectra that deviates from a normal distribution will be indicative of Faraday complexity. 

To avoid making assumptions about how complexity manifests in the spectrum, we evaluate the possible structure in the residual spectra as an additional wavelength-independent noise term $\sigma_{\rm add}$ (where `add' refers to the additional variance present), which acts to increase the scatter of the data according to
\begin{equation}
\boldsymbol{\sigma_{{\rm total}}}^{2} = \boldsymbol{\sigma}_{{\rm noise}}^{2} + \sigma_{\rm add}^{2}.
\end{equation}
This approach has been used by \citet{Allison2017} to model quasar variability and by \citet{Purcell2015} to account for poorly determined uncertainties in RM measurements. Assuming a normal distribution centred around $\mu$, the likelihood function for $\sigma_{\rm add}$ is given by
\begin{align}
  & \mathcal{L} \equiv
  {\rm Pr}(\boldsymbol{d}| \sigma_{\rm add}, \boldsymbol{\sigma}_{\rm noise}) \nonumber \\ 
  & = \frac{1}{\sqrt{(2\mathrm{\pi})^N\prod_{k}^{N}{\sigma_{{\rm total},k}^2}}}  \exp{\left[-\frac{1}{2}\sum_{k}^{N}\frac{(d_{k}-\mu)^{2}}{\sigma_{{\rm total},k}^2} \right]},
\label{eqn:like_sigma_add}
\end{align}
with $\mu=0$ expected for the residual spectrum. By Bayes' theorem, the posterior probability density ${\rm Pr}(\sigma_{\rm add}|\boldsymbol{d}, \boldsymbol{\sigma}_{\rm noise})$ can be calculated by multiplying $\mathcal{L}$ by a suitable prior function $\boldsymbol{\pi}(\sigma_{\rm add})$. Since we are ignorant of the value of $\sigma_{\rm add}$ we use a scale-invariant Jeffreys' prior, which for the noise term is:
\begin{equation}
  \pi(\sigma_{\rm add})\propto \sigma_{\rm add}^{-1}
\end{equation}
In practice we set $d(\lambda^2)=\boldsymbol{u}/\boldsymbol{\sigma_{u}}$ and $\boldsymbol{q}/\boldsymbol{\sigma}_{q}$, and $\boldsymbol{\sigma}_{\rm noise}$ to a unit vector, so that the value of $\sigma_{\rm add}$ is normalized and comparable between disparate spectra.

\begin{figure*}
  \centering
  \includegraphics[width=18.0cm, trim=0 0 0 0]{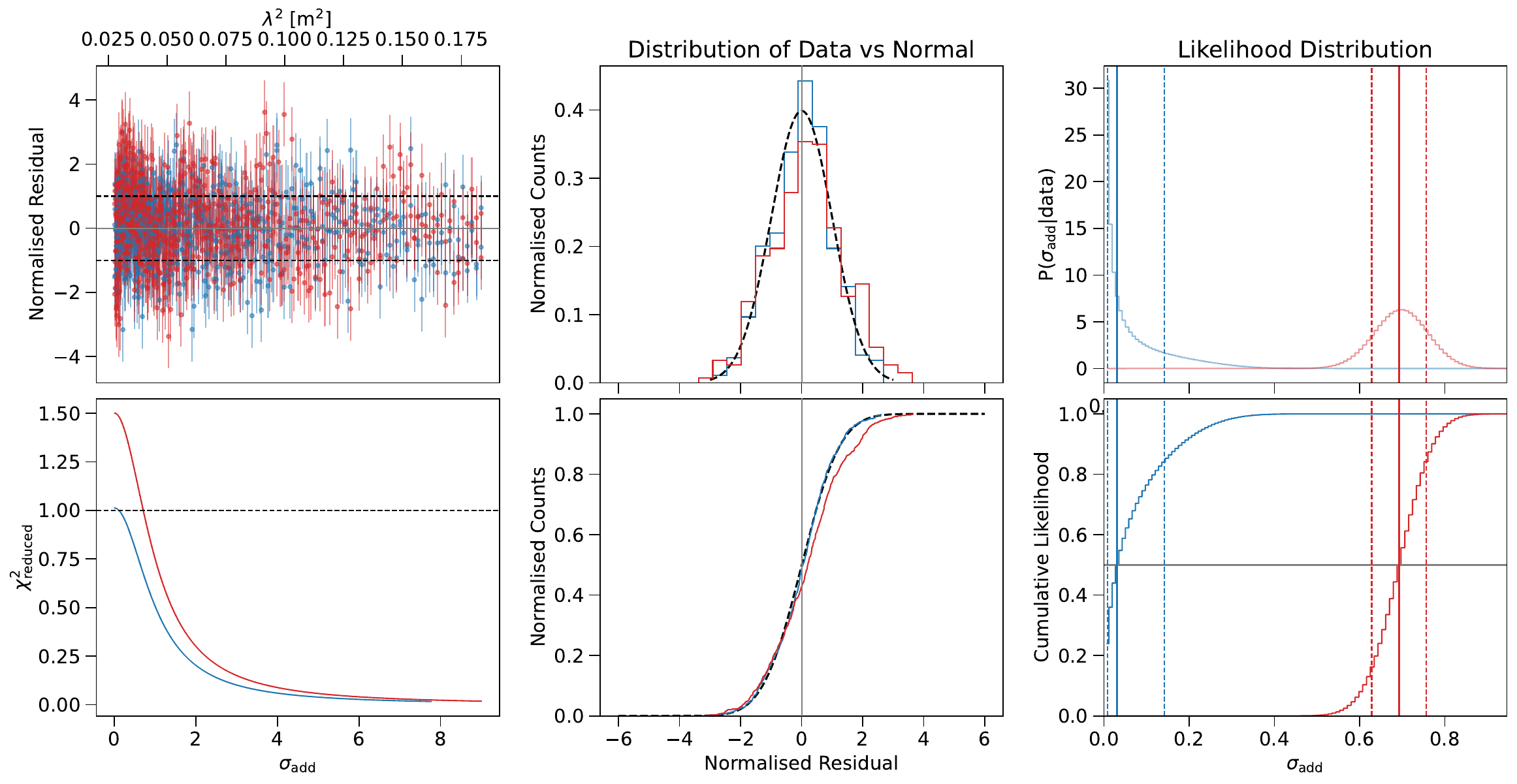}
  \caption{\small {\bf Top-left:} Residual
    $\boldsymbol{q}/\boldsymbol{\sigma}_{q}$ and
    $\boldsymbol{u}/\boldsymbol{\sigma}_{u}$ spectra of a simple (blue) and complex (red)
    source after a Faraday thin model derived from RM-synthesis has
    been subtracted. Dashed lines show the $\pm\,1\sigma$ limits. {\bf
      Bottom-left:} Reduced $\chi^2$ as a function of $\sigma_{\rm
      add}$. Values $<1$ indicate that modeling the residuals with this much
    $\sigma_{\rm add}$ would make the uncertainties
    inconsistently large compared to the data. {\bf Middle column:}
    The probability density distribution (top) and cumulative density (bottom) of the residual spectra, as compared to a normal distribution (black dashed line). A wider distribution here is indication of unmodeled structure in the residuals.
    {\bf Right
      column:} Marginal posterior probability distributions for the
    additional scatter term $\sigma_{\rm add}$.  Solid vertical lines
    show the mean of the distribution and dashed vertical lines show the $\pm\,1\sigma$ values. The simple source is best modeled with a very small $\sigma_{\rm add}$ value (consistent with zero) while the complex source requires a larger value. }
  \label{fig:residuals}
\end{figure*}

Figure~\ref{fig:residuals} shows the normalized residuals $\boldsymbol{u}_{\rm res}/\boldsymbol{\sigma}_{u}$ and $\boldsymbol{q}_{\rm res}/\boldsymbol{\sigma}_{q}$ for two simulated sources -- one Faraday-simple (blue) and one Faraday-complex (red).  We use Equation~(\ref{eqn:like_sigma_add}) to calculate the marginal posterior probability distribution over a suitable range of $\sigma_{\rm add}$ values, and report the median of the posterior distribution as the best estimate of $\sigma_{\rm add}$. The uncertainties on $\sigma_{\rm add}$ are reported as the 16$^{\rm{th}}$ and 84$^{\rm{th}}$ percentiles. In this Faraday thin case, $\sigma_{\rm add}$ is consistent with zero. For the Faraday-complex source, the distribution of values deviates significantly from Gaussian and the posterior distribution strongly constrains the additional scatter required to model the data. Because we calculate $\sigma_{\rm add}$ using the error-normalized residual, $\sigma_{\rm add}$ is a unitless quantity whose value can easily be compared between datasets.

The $\sigma_{\rm add}$ value for a source can be computed separately for Stokes $q$ and $u$, or using both parameters together. Using both Stokes parameters together is generally preferred, as it provides more statistical power to constrain the uncertainties on $\sigma_{\rm add}$. However, there can be cases where instrumental effects can be preferentially present in one Stokes parameter, which could potentially mimic Faraday complexity and bias the estimation of $\sigma_{\rm add}$. RM-Tools calculates three $\sigma_{\rm add}$ values: with Stokes $q$ only, with Stokes $u$ only, and with both together; these are referred to as $\sigma_{\rm add, q}$, $\sigma_{\rm add, u}$, and $\sigma_{\rm add_c}$ (for combined) respectively. Users who are concerned about artifacts in their data can check whether the three values show significant disagreement as a possible sign of instrumental effects associated with a single Stokes parameter.

\subsubsection{Detecting complexity in the Faraday dispersion function}
\label{sec:complexity_fdf}
Complexity in the FDF can also be assessed in the Faraday depth domain, using the FDF and assessing deviations from the ideal Faraday-thin case (as represented by the RMSF). Drawing inspiration from spectral line analysis techniques, the second moment of the FDF has been used as a measure of Faraday complexity \citep[e.g.,][]{Anderson2015, Dickey2019, Livingston2021, Erceg2022}.
The second (normalized central) moment of the FDF is defined \citep{Brown2011} as 
\begin{equation}
  M_2 = \sqrt{\frac{\sum_j^N|\tilde{\mathcal F}(\phi_{j})|\,(\phi_j-\bar{\phi})^2}{\sum_j^N|\tilde{\mathcal F}(\phi_{j})|}},
\end{equation}
where $\bar{\phi}$ is the polarization-weighted mean Faraday depth (i.e., the first moment) given by 
\begin{equation}
  \bar{\phi} = \sqrt{\frac{\sum_j^N|\tilde{\mathcal F}(\phi_{j})|\,\phi_j}{\sum_j^N|\tilde{\mathcal F}(\phi_{j})|}}.
\end{equation}

Direct calculation of any of the moments using the full FDF is strongly discouraged, as the positive-definite nature of $|\tilde{\mathcal F}|$ results in noise in the outer wings of the FDF contributing significantly to all the moments (but increasingly strongly to the higher moments). It can also be seen that changing the range of Faraday depths sampled in the FDF will also change the moments in this case. Two solutions to this problem have been used in recent publications: intensity masking, or using the clean component spectrum.
Intensity masking \citep[as used by][]{Dickey2019, Livingston2021, Erceg2022} removes the noisy parts of the FDF by including only the portions of the FDF above a user-defined threshold in the calculation of the moments. It should be noted that the resulting moments may be sensitive to the choice of threshold parameter, that applying this technique to uncleaned FDFs risks mis-identifying the sidelobes as complexity, and that the second moment will take a non-zero value even for an ideal Faraday-thin source.

It is recommended that one use the clean component spectrum produced by RM-clean (\S\ref{sec:rmclean}) instead of the (dirty or clean) FDF in the moment calculation. This removes the problem of noise contributions to the moments (provided the FDF is not significantly over-cleaned) and should produce a second moment of zero for an ideal Faraday-thin source, but similarly introduces a possible dependence on the parameters that define the cleaning performance. \citet{Anderson2015} investigated the effect of applying RM-clean to different threshold values above the measured noise, and concluded that this method is reliable for clean thresholds above a S/N level of 5.

RM-Tools implements only the clean-component based second moment complexity metric; users can straightforwardly apply intensity masking to the outputs of RM-Tools if desired.

\subsection{Additional RM-synthesis-related algorithms} \label{sec:additional_algorithms}
Several variations on RM-synthesis have been proposed \citep[e.g.,][]{Andrecut2012, Bell2012, Bell2013, Schnitzeler2015, Pratley2020, Gustafsson2025}. Two of these have been implemented in RM-Tools; the differences from conventional RM-synthesis are summarized here.
\subsubsection{Full resolution deconvolution}\label{sec:full_resolution}
\citet{Rudnick2023} presented a variation on RM-synthesis where 
the $\lambda^2_0$ term in Equations~\ref{eq:fdf} and \ref{eq:rmsf} is removed (or set to zero). This has no effect on the amplitude of the FDF, but significantly alters the phase (polarization angle) behavior of the FDF and RMSF. The desired effect is that it significantly narrows the width of the real component of the main lobe of the RMSF; if RM-clean is performed using a restoring Gaussian based on the width of the real component rather than the width of the amplitude of the RMSF, the result is a narrower restoring Gaussian (effectively a super-resolution method) and narrower peaks in the resulting clean spectrum. \citet{Rudnick2014} proposed this as a technique for improving the interpretibility of complex Faraday spectra, by improving separation of multiple components in an FDF.

This variation has been included in RM-Tools as an option under the name `super-resolution'. It sets the value of $\lambda^2_0$ to zero for the purposes of computing the FDF and RMSF, and for calculation of the intrinsic polarization angle (which, with this option, can be directly extracted from the phase of the FDF). The theoretical FWHM of the RMSF (specifically, for the real component of the RMSF) is taken from the revised equation (their Eq.~9) of \citet{Rudnick2023}. The Stokes $I$ reference frequency is kept unchanged, since that is selected internally within the fitting routine (approximately the center of the supplied frequencies) and is not rescaled to match $\lambda^2_0$. We note that only the RM-synthesis step is modified by this option; RM-clean operates identically, other than the different restoring beam width and phase behavior inside the FDF and RMSF. Due to the more rapid phase modulation present with this variation, a smaller Faraday depth step size is required to ensure that the complex FDF is properly sampled; this is handled automatically if the step sized is automatically determined, but may cause problems for users who manually specify the step size.

\subsubsection{Bandwidth depolarization correction}\label{sec:bandwidth}
\citet{Fine2023} published a modification to RM-synthesis to partially mitigate the effects of bandwidth depolarization. Bandwidth depolarization is caused by Faraday rotation across the finite bandwidth of a single frequency channel; differential Faraday rotation across the channel causes averaging of different polarization angles, producing a net loss of polarized signal. For a fixed channel width, this imposes an effective limit on sensitivity to sources with very large RMs; conversely, if an observation requires the ability to accurately measure polarization from sources with large RMs, this imposes a requirement on the bandwidth of individual channels. Bandwidth depolarization is effectively applied at the time of observation, and cannot be fully reversed in post-processing. \citet{Fine2023} proposed two modifications to the RM-synthesis transform (Eq.~\ref{eq:fdf}) to improve detection of highly bandwidth depolarized sources and to correctly characterize their polarized properties (at the cost of increased noise), and developed a two-step algorithm that first searches for significant polarized peaks using the first modification, then characterizes the strongest peak using the second modification. This algorithm is not compatible with RM-clean (although an adapted version of RM-clean could be possible).

This modified algorithm has been implemented as an alternative tool within RM-Tools, which a user can choose to use instead of the default RM-synthesis tool. The run time of this alternative is typically 2-5 times longer than the default, but otherwise it takes effectively the same inputs and produces similar outputs. An additional tool, which predicts the potential strength of bandwidth depolarization as a function of RM for a given channel configuration, has also been included to allow users to determine the RM range in which a given dataset may be significantly affected.

\subsection{QU-fitting}\label{sec:QUfitting}

RM-Tools' QU-fitting implementation uses nested sampling \citep{Skilling2004} for the fitting process. For flexibility, RM-Tools uses the Python package {\it bilby} \citep{bilby_paper}, which provides an interface that can use any of several nested sampling packages implemented in Python; at runtime the user can select which nested sampler to use. Of the samplers currently available, we have tested QU-fitting with the nested sampling packages {\it pymultinest} \citep{Buchner2014}, {\it nestle}\footnote{https://github.com/kbarbary/nestle}, and {\it dynesty} \citep{Speagle2020, Koposov2024}; the results are presented in Appendix~\ref{sec:sampler_qufit}.

When fitting models containing a linear combination of identical sub-models as described in $\S$\ref{sec:clumps}, models are subject to a ``labelling degeneracy''; that is, it does not matter which component we call $\mathcal{P}_1$, or $\mathcal{P}_2$, and so on. This can result in the fitting algorithm identifying multiple solutions (corresponding to different permutations of how model components are assigned to data features) and then failing to select a single representative solution (e.g., some packages compute the mean or median of the likelihood function for each model parameter).
To avoid these problems, we have implemented constraints in the models that force the fitter to consider only a single permutation. The constraint we have used is to require that the RM of the first component to be smaller than the RM of the 2nd component (and likewise for the third component, where applicable); this forces the algorithm to only consider the parts of the parameter space where the components are ordered from least to greatest RM. We have also implemented a constraint that the sum of the polarized fractions must be less than or equal to 1 (preventing models that contain too much total polarized signal), but this constraint is optional and can be removed if a user finds it unhelpful.

RM-Tools comes with several pre-made models, described in Appendix~\ref{sec:pol_models} and Table~\ref{tab:models}, which cover the most common physical models used in the literature to-date. Users can define their own models fairly straightforwardly: a model requires a function that calculates the complex polarization for an input parameter dictionary and vector of $\lambda^2_i$ values, a list of model parameters with priors and maximum/minimum values, and any constraints that should apply. The existing models demonstrate the form and functionality, and can be adapted as needed by users.

\section{RM-Tools - User interface} \label{sec:interface}
RM-Tools contains five top-level scripts, covering the most important and commonly used features of the package, as well as a number of helper scripts; these are designed to be used through a standard terminal interface and most are set up to be importable into user scripts or interactive environments (e.g., {\it Jupyter} notebooks). The top level scripts are divided by whether they operate on one-dimensional Stokes~{\it I}, {\it Q} and~{\it U} spectra, or on three-dimensional position-position-frequency Stokes {\it IQU} data cubes (which have the same analysis performed pixel-by-pixel). The underlying functionality is, where practical, the same for both data formats, except for QU-fitting which is too computationally intensive for a 3D implementation.

All the scripts within RM-Tools have a few common properties that merit remarking on. In all cases, the expected input files are in either the `Flexible Image Transport System' (FITS, \citealt{Wells1981}) format or unformatted text files. All of the scripts should be safe against NaNs in the input data, allowing the use of NaN values to denote missing or removed data.

These scripts are summarized here along with some tips for using them effectively; full information on their use can be found by invoking the `\texttt{-h}' flag in the terminal, looking at the doc strings for the top-level functions, or in the GitHub wiki\footnote{https://github.com/CIRADA-Tools/RM-Tools/wiki}.

\subsection{rmsynth1d}\label{sec:rmsynth1d}

The \texttt{rmsynth1d} script performs RM-synthesis on a single polarized spectrum and analyzes the brightest peak in the resulting FDF. It takes as input, if run from the command line, a single ASCII text file containing (whitespace-separated) columns with the channel frequencies, Stokes $IQU$ channel values, and uncertainties in those channel values. The process of extracting the spectrum is left to the user; the script is agnostic to the specific type or unit of spectrum provided (e.g., peak pixel vs. source-integrated, intensity vs. brightness temperature). 
The uncertainties are technically optional, in that a user can set them to an arbitrary constant; they affect the channel weighting for RM-synthesis and the calculation of uncertainties in all derived quantities. For valid output uncertainties, the input uncertainties should reflect the 1-sigma uncertainties in the channel Stokes values.

\texttt{rmsynth1d} performs four main steps: Stokes $I$ fitting and normalization (\S \ref{sec:model_I}), RM-synthesis (\S\ref{sec:rmsynth}), characterization of brightest FDF peak (\S\ref{sec:properties}), and Faraday complexity estimation (calculating $\sigma_{\rm add}$, \S\ref{sec:complexity_qu}). Depending on which command line flags are set, three types of output are possible: a summary of the polarization properties can be printed to the terminal, interactive plots of the spectra and FDF are shown to the user, and/or the resulting arrays (FDF and RMSF) and output properties are saved to files.

\subsection{rmclean1d}\label{sec:rmclean1d}

The \texttt{rmclean1d} script performs RM-clean (\S \ref{sec:rmclean}) on a single FDF, then analyzes the brightest peak of the cleaned and restored FDF (\S\ref{sec:properties}). It takes as inputs the products of the \texttt{rmsynth1d} script, and it has options to output the clean component FDF, the cleaned and restored FDF, and a list of the polarization properties derived using the cleaned FDF, as well as plots of the clean components and cleaned FDF. The parameters of the clean algorithm (initial clean threshold, windowed threshold, gain, and maximum iterations) can all be specified by the user on the command line.

\subsection{rmsynth3d}\label{sec:rmsynth3d}

The \texttt{rmsynth3d} script performs RM-synthesis for each line of sight (pixel) within an image. By default it performs {\it only} RM-synthesis -- the other steps that are included in \texttt{rmsynth1d} are separated off into other tools, to distribute the computational cost -- these will be introduced in a later subsection (\S \ref{sec:othertools}). \texttt{rmsynth3d} takes as required inputs 3 files: Stokes $Q$ and $U$ FITS cubes, and a text file with a list of channel frequencies. The input FITS cubes are required to have 2 or three non-degenerate axes: a frequency axis (iterating over all the channels to be used for RM-synthesis), and one or two position axes; all additional axes (including the Stokes axis) must be degenerate (axis length of 1).

Depending on the size of the input data products and the RM-synthesis parameters selected, the computational overhead of \texttt{rmsynth3d} can be significant in terms of CPU use, RAM required, and output file sizes. The memory footprint in particular can be quite restrictive: \texttt{rmsynth3d} requires the input data (Stokes $Q$ and $U$, and optionally $I$) be loaded along with the complex FDF cube, which can result in memory requirements in excess of 100 GB if large ($>$10 GB) cubes are used -- overwhelming the capacities of most computational systems. Users are encouraged to carefully decide the parameters that set the number of samples in the FDF (the maximum Faraday depth and either the Faraday depth step size or the number of samples across the FWHM of the RMSF) in order to optimize the size of the resulting FDF cubes. If the required memory exceeds the capacity of the computing system, users are recommended to break the data into smaller pieces (such as using the \texttt{createchunks} tool in \S \ref{sec:othertools}).

No CPU parallelization has been developed for \texttt{rmsynth3d}. During testing, we found that the actual RM-synthesis calculations are very light-weight in terms of CPU operations per MB of input data. In nearly all cases, the run time was dominated by the time spent reading and writing the data to disk, so we elected not to spend development time optimizing the (short) CPU-bound phase of the script.

The primary outputs of \texttt{rmsynth3d} are the FDF and RMSF cubes; for each of the FDF and RMSF, three cubes are stored: the real and imaginary components (corresponding to Stokes $Q$ and $U$ respectively) and the `total' or polarized intensity. The volume of the output cubes is typically somewhat larger than the input cubes, as oversampling of the FDF often results in more Faraday depth samples than original frequency channels. In cases where there is no position-dependence of the missing channels (i.e., they are uniformly blanked across the whole image plane) and no position-dependent weights are applied (i.e., Stokes I normalization), the RMSF has no position-dependence; the full RMSF cube is still generated (for compatibility with \texttt{rmclean3d} and other tools), but the resulting files are extremely compressible with standard file compression methods (e.g., tar, zip). We have observed use cases where 2.5 GB RMSF cubes can be compressed down to 37 MB for archiving. If data volume is a problem for long-term storage/archiving, the `total' (amplitude) cubes can be removed, as they can be regenerated from the real and imaginary cubes.

In addition to the cube products, three images are produced: the FWHM of the RMSF, the maximum polarized intensity of each FDF, and the RM corresponding to that maximum. The last two are determined on a per-pixel basis, but using only the sampled values of the FDF (rather than peak-fitting, as is done in \texttt{rmsynth1d} and \texttt{peakfitcube}).

\subsection{rmclean3d}

Just as the \texttt{rmclean1d} script applies RM-clean to the outputs of \texttt{rmsynth1d}, the \texttt{rmclean3d} script performs the same function on the outputs of \texttt{rmsynth3d}. This script takes as inputs the FDF and RMSF cubes produced by \texttt{rmsynth3d}, and applies RM-clean independently to each individual FDF. The outputs are output cubes for the clean components (real, imaginary, and `total') and the cleaned and restored FDF cubes (a similar set of 3 files).

Due to the higher computational cost of the RM-clean algorithm (compared to RM-synthesis), \texttt{rmclean3d} supports CPU parallelization, distributing groups of FDFs among separate threads to speed up computation. This parallelization is mediated by the Python module {\it Schwimmbad} \citep{schwimmbad}, and supports both Message Passing Interface (MPI) and standard Python multiprocessing.

\subsection{qufit}

The \texttt{qufit} script provides a command line interface to the QU-fitting module. It is designed to work on single FDFs (similar to the 1D RM-synthesis and -clean scripts); no 3D QU-fitting tools have yet been developed due to the expected prohibitive computational cost. This script takes as an input the same form of ASCII file as \texttt{rmsynth1d}. Stokes $I$ fitting and normalization is performed, and the resulting fractional Stokes spectra are fit by a user-selected QU model using a nested sampling algorithm. The script produces several outputs: text files storing the fitted model parameters (with uncertainties) and goodness of fit metrics, a standard `corner plot' showing the posterior parameter distributions, and a figure showing the spectra and fit.

RM-Tools comes with a number of pre-made QU-fitting models, which users can fit as-is or use as examples for creating new models; these are listed in Table~\ref{tab:models}. Users should note that, depending on the observational setup and science goals, the sampled parameter space of these pre-made models may not suit their specific needs. These can be easily adjusted within the model files.

The QU-fitting module has several ways in which its operation is very sensitive to user input. Unlike \texttt{rmsynth1d}, where the channel uncertainties are only needed to produce accurate uncertainties in the derived parameters, QU-fitting requires accurate channel uncertainties for effective convergence of the fit. Channel uncertainties that are too small (i.e., smaller than the actual noise in the data) prevent the fitting routine from recognizing when it has found a good fit, while uncertainties that are too large cause it to halt fitting potentially too early and lead to inaccurate uncertainties in the model parameters.

\texttt{qufit} is also sensitive to the choice of nested sampling module. Our script has been built using the Python module {\it bilby}, which allows users to select from several different Python nested sampling implementations. In testing, we have observed that different samplers (e.g., {\it dynesty, pymultinest, nestle}) have different behaviours in terms of speed and reliability of convergence. The results of our basic tests can be found in Appendix~\ref{sec:sampler_qufit}; based on those tests we found the {\it dynesty} or {\it nestle} samplers worked best in the cases we considered.

The complexity of the model and the size of the parameter space have also been found to significantly affect the probability and rate of convergence of the fitting. More complex models take longer to fit and have a higher chance of failing to converge, particularly those with 2 or more separate polarized components. It has been observed that in some cases the script will continue to run indefinitely until overflowing available memory; this appears to be connected to the initial conditions of the nested sampling, as re-running the same spectrum usually results in convergence within the expected time. The size of the parameter space affects the convergence time, particularly the size of the RM search range. The run time of the script can be significantly improved if users are able to limit the parameter range based on assumptions about the source properties (e.g., estimating the expected range of RMs for a given source's region of sky); this can be done by modifying the model files, which contain the limits of the parameter search space.

Since \texttt{qufit} operates on a single spectrum and model per execution, it does not provide any internal capability to compare model fits and to select the best fitting model. It is left to the user to decide how they will perform this task. A detailed discussion of model selection is left to a forthcoming analysis of QU-fitting effectiveness (L.~Oberhelman et al., submitted for publication).

\subsection{Additional tools}\label{sec:othertools}
In addition to the main scripts described above, several additional command-line scripts are included in RM-Tools to provide both convenience functions for common tasks and easier access to some internal functions of RM-Tools. Each of these are prefixed with \texttt{`rmtools\_'} on the command-line.

\subsubsection{calcRMSF}
The \texttt{calcRMSF} script provides convenient access to the RMSF calculation functions for the purposes of predicting the behavior of the RMSF for different channel / frequency configurations. It takes as input either a file with channel frequencies (and optional channel weights) or command-line flags that define a lowest frequency, highest frequency, and channel width (in which case it assumed that the frequency coverage is uniform across that range). The script calculates the RMSF corresponding to this frequency coverage and generates a figure with the RMSF and gives several key properties of the RMSF (e.g., main lobe FWHM, sensitivity to broad scales, sidelobe level).

\subsubsection{freqfile}
The \texttt{freqfile} script takes as input a FITS cube with a frequency axis, and generates a text file containing the list of channel frequencies. This output file can be used directly with \texttt{rmsynth3d} or \texttt{calcRMSF}.

\subsubsection{fitIcube}
The \texttt{fitIcube} script performs Stokes $I$ spectral fitting pixel-wise to a supplied Stokes $I$ FITS cube. Each pixel is fit fully independently, without enforcement of smoothness of fit parameters across adjacent pixels. The script first determines the noise per-channel, by computing the MADFM, removing pixels more than 5 sigma away from the median (which are assumed to be pixels containing radio sources), and recomputing the MADFM with the remaining pixels; this channel noise is required to apply channel weighting to the fit and to properly determine uncertainties on the fit parameters. If the user selects it, a pixel mask can be calculated, identifying pixels with at least one channel above either a user-supplied intensity threshold or a user-selected multiple of the noise level. If the mask is used, only pixels in the mask are fit and all other pixels have NaN values in the outputs.

The script then performs the spectral fitting (\S \ref{sec:model_I}); this is parallelized using Python's {\it multiprocessing} module to improve computational speed. Three sets of outputs are produced: a pixel mask (if used) as a 2D FITS image, a series of 2D FITS images for all the fit parameters and their corresponding uncertainties (as well as the full covariance matrix), and a 3D FITS cube containing the Stokes $I$ model values for each channel. This model cube is suitable for use in \texttt{rmsynth3d} if desired.

If the default model (log-polynomial) is used, the zeroth coefficient map corresponds to the intensity at the reference frequency, while the first coefficient map is the spectral index.

\subsubsection{peakfitcube}
The \texttt{peakfitcube} script performs polarization characterization (\S \ref{sec:properties}) pixel-wise to a pair of 3D FDF cubes (containing the real and imaginary components of the FDF), such as that produced by \texttt{rmsynth3d} or \texttt{rmclean3d}. The inputs are these FDF cubes, a file with the channel frequencies, and an optional file with the channel noise values (required to calculate uncertainties). The outputs are a series of 2D FITS images of the polarized properties characterized from the FDF.

\subsubsection{createchunks and assemblechunks}
The \texttt{createchunks} and \texttt{assemblechunks} scripts were made to break large FITS cubes into smaller pieces that can be processed by the 3D tools (\texttt{rmsynth3d}, \texttt{rmclean3d}, \texttt{fitIcube}, and \texttt{peakfitcube}) without overflowing available memory, then reassemble the outputs. The \texttt{createchunks} script takes an input cube and divides it into many smaller files with a user-specified number of pixels; this is done in a way that removes the position information and reduces the outputs to two dimensions (frequency and pixel index). These `chunks' can be processed by the other 3D tools, which will produce outputs with similar dimensions. The \texttt{assemblechunks} script takes these output products and reassembles them to have the same image-plane dimensionality as the original input cubes.

The \texttt{createchunks} script is very memory efficient -- its memory footprint is not significantly larger than the size of a single chunk. The \texttt{assemblechunks} module must hold an entire output file in memory at one time; this is no problem for 2D products such as the maps from \texttt{peakfitcube}, but can be quite large for reassembling FDF or RMSF cubes.

\subsubsection{3DIrescale}
The \texttt{3DIrescale} tool converts Stokes $I$ model parameters to a new reference frequency (normally the reference frequency for the polarization results). The Stokes $I$ model fitting and the RM-synthesis generally use different reference frequencies -- when Stokes $I$ normalization is performed, the fitting must be done before the RM-synthesis, and the results of the fitting can change the resulting channel weights (and thus the optimum $\lambda^2_0$ value used). It can be inconvenient to have the Stokes $I$ and polarization properties referenced to different frequencies, so RM-Tools has the capacity to translate the model parameters to match the polarization reference frequency. In \texttt{rmsynth1d} this is done automatically, but in the 3D case the fitting and RM-synthesis steps are separated so a standalone tool is needed to make the correction. This script takes as input the model parameters and covariance matrix maps, and a user-supplied reference frequency or $\lambda^2_0$ value; it outputs new parameter and uncertainty maps. Note that the new parameter uncertainties are generated by first-order Taylor-approximation of the model conversion equations, so the new uncertainties are likely not reliable if the reference frequency is shifted by more than approximately 10\% (at which point a warning is raised).

\subsubsection{bwdepol and bwpredict}
The \texttt{bwdepol} and \texttt{bwpredict} scripts provide implementations of the bandwidth depolarization correction version of RM-synthesis (\S \ref{sec:bandwidth}). \texttt{bwdepol} is a variation of \texttt{rmsynth1d} that applies the bandwidth depolarization correction of \citet{Fine2023}; it takes the same inputs and outputs as \texttt{rmsynth1d}, with the exception that the input file may optionally have an additional column that gives the channel bandwidths. If this is not supplied, the channel bandwidth is assumed to be the spacing between the first two channels in the file, and is assumed to be equal for all channels. The output includes additional plots that characterize the strength of the bandwidth depolarization at the RM of the strongest peak.

The \texttt{bwpredict} script generates figures that show the strength of bandwidth depolarization as a function of RM for a user-supplied frequency configuration. The input is a file of channel frequencies (and, optionally, channel widths). This tool is intended for use when planning observations or data reduction, to decide what frequency resolution is needed to prevent bandwidth depolarization from being a large problem for later analysis.

\section{Testing of software performance}\label{sec:tests}
In the following subsections we summarize tests performed to verify the accuracy and effectiveness of the key new features offered by the RM-Tools package.

\subsection{Testing the Complexity Metrics}
We have tested the ability of our proposed metrics to detect complexity using two types of model: 1) a uniform slab (Eqn.~\ref{eqn:slab}) and 2) a turbulent foreground screen model (Eqn.~\ref{eqn:Burn_law}). The degree of complexity in both models depends on only a single parameter, allowing us to assess the performance of our methods over a range of S/N. 

To test various complexity metrics, we constructed simulated polarized spectra by selecting a frequency range and channel width (for the tests discussed below we used 1-2 GHz with 10 MHz channels, but other configurations representative of major radio telescopes were also explored with similar results), choosing an array of test values for S/N and the width parameter ($\Delta\phi$ for the uniform slab model, $\sigma_{RM}$ for the turbulent screen model), generating model $Q$ and $U$ spectra given those model parameters, and adding noise to the $Q$ and $U$ spectra. The noise amplitude was selected to produce the targeted S/N of the peak in the resulting FDF. The resulting spectra were processed through RM-synthesis and RM-clean and the complexity metrics were extracted from the outputs. Both the $\sigma_{add,C}$ and $M_2$ metrics were analyzed.

\begin{figure*}
  \centering
  \includegraphics[width=\textwidth, trim=0 0 0 0]{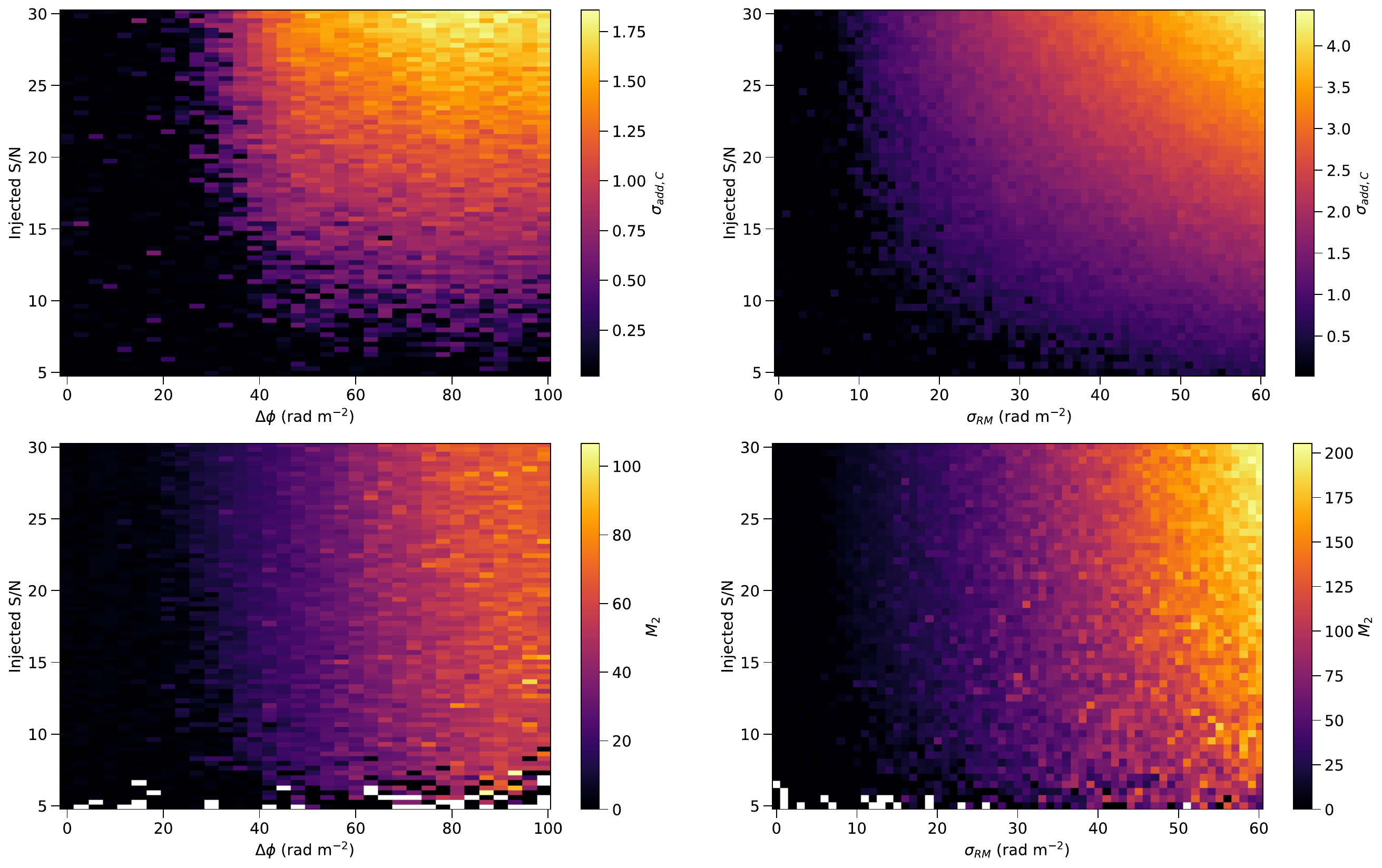}
  \caption{\small Complexity metric values resulting from the uniform slab (left) or turbulent screen (right) models for different S/N and Faraday thickness values. The complexity metrics shown are $\sigma_{add,C}$ (top row) and $M_2$ (bottom row). Each pixel was derived from a single polarized spectrum realization, so inter-pixel variations indicate the influence of spectral noise on the resulting metric value. Blank (white) values in the $M_2$ plots occur occasionally at low S/N when the resulting FDF contains no samples above the CLEAN threshold, and thus no clean components with which to compute the second moment.}
  \label{fig:complexity_tests}
\end{figure*}

Figure~\ref{fig:complexity_tests} shows an example of these tests; each panel shows a complexity metric value as a function of S/N and Faraday thickness, for a single realization of a random polarized spectrum. An ideal metric would show increasing strength as a function of Faraday thickness, would be as independent as possible of S/N (i.e., would be represented as gradient from left to right), would be comparatively insensitive to the noise in the $Q$ and $U$ spectra (represented in the figure as `noisy' variability across adjacent pixels), and would function equally well for any form of Faraday complexity (i.e., different models). In Fig.~\ref{fig:complexity_tests} we can see that neither of the metrics are ideal, and that each has different strengths and weaknesses. 

The $\sigma_{add}$ metric shows the strongest dependence on S/N (with faint spectra never showing complexity), but has the lowest sensitivity to noise. It also shows signs of saturating, in that the value does not increase beyond some value even as the Faraday thickness increases. The noise sensitivity and saturation can be partially mitigated by normalizing $\sigma_{add}$ by its reported uncertainty -- the resulting parameter has less variation due to noise and continues increasing in value as Faraday thickness increases.

The $M_2$ metric, by comparison, has weaker dependence on the S/N and a more consistent increase in value with increasing Faraday thickness. It is slightly more sensitive to noise (especially for the turbulent screen model). It also comes with the problem of being parameter dependent: its value can change significantly with the clean threshold chosen. A deeper clean threshold produces larger and noiser values for $M_2$; larger values are smaller and less noisy, but also are more likely to have undefined values at low S/N (in cases where no clean components found about the selected threshold).

\citet{Thomson2023} and \citet{Vanderwoude2024}  performed comparisons of the $\sigma_{add}$ and $M_2$ metrics for populations of polarized sources observed in the SPICE-RACS and POSSUM Pilot 1 surveys, respectively. Both found that the two metrics are correlated, and that thresholds in each can be identified where both metrics generally (but not always) agree if a source is simple or complex. They also observed that sources with higher signal-to-noise ratio in polarization are more likely to be classified as complex, possibly due to the complexity being masked by noise in weakly detected sources.

\subsection{Testing accuracy and uncertainty estimates with simulations}
\label{sec:errortests}
We performed a series of tests with simulations to evaluate the correctness of the reported measurements and the corresponding uncertainty estimates reported by the RM-synthesis/RM-{\scriptsize CLEAN} and QU-fitting tools. The simulations were performed by generating synthetic Stokes spectra for a single Faraday-thin component (with randomly selected Faraday depth and initial polarization angle, and fixed fractional polarization) and adding Gaussian random noise with a fixed variance to both Stokes $Q$ and $U$, then running these spectra through the 1D RM-synthesis and QU-fitting tools and comparing the measured results against the ground truth of the simulations.

In the results that are presented below, we used a frequency coverage matching the POSSUM Band 1 survey: 288 channels of 1 MHz from 800 to 1088 MHz (other frequency ranges were tested and found to give consistent results). For the main series of tests, the Stokes $I$ spectrum was not simulated and a flat spectrum was assumed, and all channels were assigned the same noise variance. For each test, 10000 iterations were performed for the RM-synthesis tool and 1000 iterations were performed for the QU-fitting tool (due to the higher computational cost). A range of signal to noise (in the FDF) from 8 to 100 was explored and found to give the same results; the results presented below are for a signal to noise ratio of 17 (signal to noise of 1 per channel).

\begin{figure*}
  \centering
  \includegraphics[width=15.5cm]{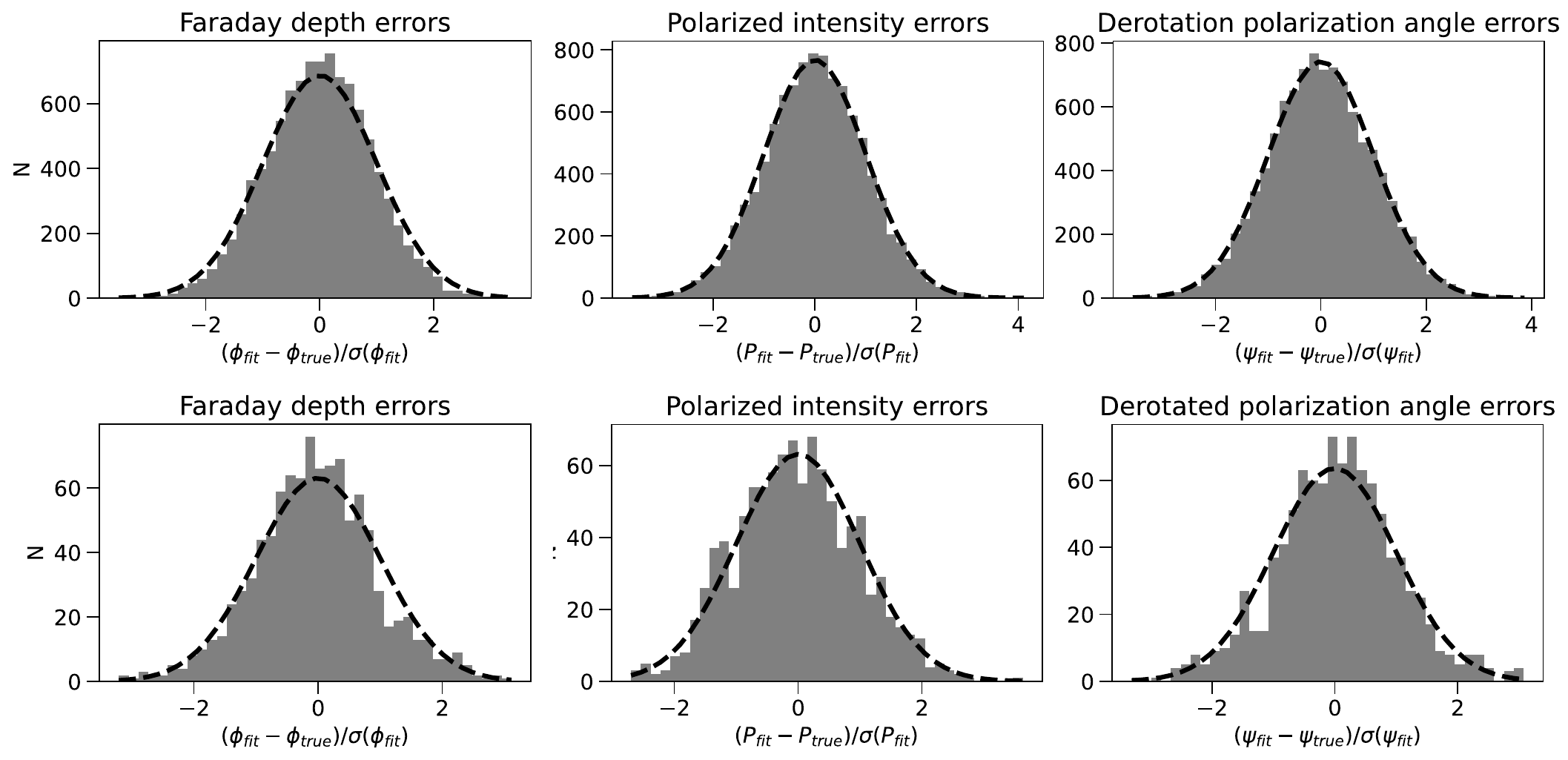}
  \caption{\small Error distributions (normalized by reported uncertainty) in the Faraday depth (left), polarized intensity (center), and de-rotated polarization angle (right), for RM-synthesis (top row) and QU-fitting (bottom row). All distributions appear well described by a Gaussian of unit width (dashed lines)}
  \label{fig:uncertainty_histograms}
\end{figure*}

The key quantities that we tested were the fitted peak Faraday depth  $\phi_{\rm fit}$ and its uncertainty, the fitted polarized intensity $P_{\rm fit,eff}$ and its uncertainty, and the de-rotated polarization position angle $\psi_{\rm fit}$ and its uncertainty.
For each of these quantities, we computed the difference between the measured values and the ground truth values, normalized by the reported uncertainty, for all iterations and then looked at the resulting distribution; if the reported uncertainties accurately describe the residual errors then this distribution should have a standard deviation of 1.
Samples of these distributions appear in Figure \ref{fig:uncertainty_histograms}, with a Gaussian distribution of standard deviation 1 overlaid. 
The residual distributions were found to be well matched with a Gaussian and had a standard deviation close to the expected value; the exact values are shown in Table \ref{tab:error_stddevs}. In all cases, the mean of the distributions was consistent with zero, indicating no bias in the reported values.
The estimated uncertainties on the values in the table (i.e., the uncertainty on the degree to which our reported uncertainties are accurate) are 0.007 for the RM-synthesis and 0.022 for the QU-fitting. The uncertainties in the Faraday depth appear to be overestimated by approximately 7\% with RM-synthesis and 2.5\% with QU-fitting, but the other uncertainties are accurate to within 2\%. We have not been able to identify a specific cause of the overestimated Faraday depth uncertainties, but we did find indications that the degree of overestimation may vary slightly with different frequency coverage (ranging from 5\% to 10\% in the tests that we performed).

\begin{table}
\begin{center}
    
\begin{tabular}{|c|l|l|}
\tableline
Parameter     & rmsynth1d & qufit  \\
\tableline
$\phi_{\rm fit}$ & 0.934 & 0.976 \\
$P_{\rm fit}$     & 1.001 & 0.979 \\
$\psi_{\rm fit}$     & 1.003 & 0.988 \\
\tableline
\end{tabular}
\end{center}

\caption{Measured standard deviations in the measured residuals (normalized by estimated uncertainty) for key quantities. Correctly estimated errors correspond to a value of 1, overestimated errors values less than 1, and underestimated errors values greater than 1. \label{tab:error_stddevs}}
\end{table}

We note that these results are only demonstrative for the ideal case of a Faraday-thin source and Gaussian random noise; a detailed treatment of uncertainties in the presence of arbitrary Faraday complexity or instrumental artifacts is beyond the scope of this work.

\section{Summary and discussion}\label{sec:summary}

The analysis of radio polarization data has, with the development of new methods such as RM-synthesis and QU-fitting, become more algorithmically complex to the level where it is difficult for most researchers to develop their own software tools for the task. We have developed RM-Tools, a powerful, flexible, accurate, and feature-rich Python-based open-source toolkit for radio polarization analysis, with the intention of making such analysis easier, faster, higher quality, and more uniform. We have released this toolkit as a standard Python package, downloadable via GitHub\footnote{https://github.com/CIRADA-Tools/RM-Tools}, the Python Package Index (PyPI)\footnote{Installation via PyPI can be performed in the standard way using the Python `pip' module as `pip install RM-Tools'.}, and Zenodo (doi:10.5281/zenodo.17851649)\footnote{We have uploaded the version at time of publication (v1.4.10) to Zenodo (doi:10.5281/zenodo.17851650), and intend to upload all major versions henceforth.}. RM-Tools was released under the MIT open-source license.

RM-Tools is split into two sub-modules; the first is focused on the analysis of individual polarized spectra (`1D'), and the second is focused on analysis of spectral data cubes (`3D'). Two techniques for polarized analysis are included: RM-synthesis with (optional) RM-clean, and QU-fitting (only available for 1D due to high computational cost). Additional features of the package include:
\begin{enumerate}
    \item Stokes $I$ spectral modeling, using polynomial or log-polynomial models, with fixed or dynamic order selection
    \item normalization of polarized spectra by Stokes $I$ models
    \item calculation of the $\sigma_\mathrm{add}$ and $M_2$ complexity metrics
    \item characterization of properties of the highest peak in the FDF
    \item uncertainty estimation on derived polarization properties
    \item full-resolution deconvolution as described by \citet{Rudnick2023}
    \item bandwidth depolarization prediction and correction as described by \citet{Fine2023}
    \item `windowed' deep cleaning of FDFs
    \item characterization of the RMSF properties for arbitrary frequency configurations
    \item breaking large data cubes into manageable pieces, and reassembling the resulting RM-synthesis products
    \item CPU-based parallelization of computationally expensive 3D operations (RM-clean, spectral modeling, peak fitting)
    \item ten different QU-models for fitting, with the ability of users to define custom models straightforwardly
    \item command-line and Python-importable interfaces to all main features.
\end{enumerate}

With RM-Tools we have proposed a new metric to assess the presence of Faraday complexity in a polarized spectrum, in the form of $\sigma_\mathrm{add}$. This metric estimates how much additional variance is required to explain the observed signal beyond what is predicted by a Faraday simple model with noise. While this metric does not provide a measure of the Faraday thickness of a complex source (in contrast with the $M_2$ metric), it is less sensitive to noise or variations in RM-clean parameters, as well as being less computationally expensive (in that it can operate on dirty FDFs without using RM-clean).

Since the first official release of RM-Tools in May 2020, it has accrued 70 citations reported by the Astrophysics Data System (ADS) at the time of writing.\footnote{https://ui.adsabs.harvard.edu/abs/2020ascl.soft05003P/citations} While part of the popularity can be explained by its internal use by the pipelines of POSSUM and adoption by many members of the POSSUM collaboration, analysis of the citations shows that it has been adopted by a much broader cross-section of radio astronomy, such as the pulsar \citep[e.g.,][]{Miao2023} and fast radio burst \citep[e.g.,][]{Pandhi2025} communities. We present this as evidence of the significant value generated by the development of substantial software toolkits designed for use beyond the needs of individual studies or projects. RM-Tools has benefited greatly from community engagement with the package, in the form of code contributions, feature requests, and bug reports. We request that users kindly cite this article when publishing work done using this code, and optionally the Zenodo DOI for the major version they used.

Active development of RM-Tools has ended for the present time due to limited human resources; the package is being actively maintained as bug reports or changes in dependencies are found. Since the current maintainers are early-career researchers, we are actively developing a plan to ensure continuity of maintenance for as long as can be supported. External contributions to the package are welcomed; users interested in developing new functionality are encouraged to contact the maintainers to discuss integration into the package. Bug reports and feature requests are handled primarily through the GitHub repository.

While there are currently no formal plans for a successor or major overhaul for RM-Tools, the broad community use has revealed limitations in the current design. The most significant limitation, which will become significantly worse as data volumes continue to increase, is the memory footprint required for the 3D tools. The current 3D RM synthesis implementation is most strongly affected; it requires both the input Stokes $Q$ and $U$ cubes as well as the complex FDF cube in memory at the same time. The `chunk' tools are an inefficient work-around, as they require two additional read-write steps for the full data. We have envisioned alternative implementations that require holding only one cube in memory (frequency-Stokes or FDF), but these would require a substantial redesign of the core functions of RM-Tools (for which the development resources are currently not available). We expect that the data volumes expected from Square Kilometre Array data products will drive the need for the development of the next iteration of radio polarization toolkit, and that this development will focus on improvements to computational footprint and scalability.

\begin{acknowledgments}
RM-Tools is dependent on the following scientific software packages: \texttt{astropy} \citep{astropy:2013, astropy:2018, astropy:2022}, \texttt{matplotlib} \citep{Hunter:2007}, \texttt{numpy} \citep{numpy}, \texttt{python} \citep{python}, \texttt{scipy} \citep{2020SciPy-NMeth, scipy_14880408}, \texttt{Bilby} \citep{bilby_paper, Bilby_2602178}, \texttt{corner.py} \citep{corner-Foreman-Mackey-2016, corner.py_14209694}, and \texttt{schwimmbad} \citep{schwimmbad}.

This work has also made use of \texttt{pymultinest} \citep{Buchner2014}, \texttt{nestle}\footnote{https://github.com/kbarbary/nestle}, \texttt{dynesty} \citep{Speagle2020, Koposov2024}, and \texttt{Jupyter} \citep{2007CSE.....9c..21P, kluyver2016jupyter}.

Software citation information aggregated using \texttt{\href{https://www.tomwagg.com/software-citation-station/}{The Software Citation Station}} \citep{software-citation-station-paper, software-citation-station-zenodo}.

CRP, BMG and XHS were supported by the Australian Research Council through grant
FL100100114. Parts of this research were conducted by the Australian
Research Council Centre of Excellence for All-sky Astrophysics
(CAASTRO), through project number CE110001020. The Dunlap Institute is funded through an endowment established by the David Dunlap family and the University of Toronto. CLVE, JLW and BMG acknowledge the support of the Natural Sciences and Engineering Research Council of Canada (NSERC) through grant RGPIN-2015-05948, and of the Canada Research Chairs program. The Canadian Initiative for Radio Astronomy Data Analysis (CIRADA) is funded by a grant from the Canada Foundation for Innovation 2017 Innovation Fund (Project 35999) and by the Provinces of Ontario, British Columbia, Alberta, Manitoba and Quebec, in collaboration with the National Research Council of Canada, the US National Radio Astronomy Observatory and Australia’s Commonwealth Scientific and Industrial Research Organisation. 
SPO acknowledges support from the Comunidad de Madrid Atracción de Talento program via grant 2022-T1/TIC-23797, and grant PID2023-146372OB-I00 funded by MICIU/AEI/10.13039/501100011033 and by ERDF, EU.
\end{acknowledgments}


\begin{contribution}
Author CRP developed the core functionality of the original RM-Tools code, along with an initial draft of this manuscript. CLVE led the conversion to a publicly-usable package and the subsequent writing of this manuscript. Authors LB through JLW contributed to the development of the code through feature development and/or bug fixes. Authors SI through TW contributed to testing, bug reports, or during early development. Authors TA through XZ contributed advice during development, comments on the manuscript, and general participation in the POSSUM collaboration. Authors CRP and BMG conceived the original concept of RM-Tools and planned out the initial functionality, using funding obtained by BMG.

\end{contribution}

\bibliography{reference_list}{}
\bibliographystyle{aasjournalv7}

\appendix
\section{Models of polarized spectra}\label{sec:pol_models} 


Observed structures are often much more complicated than the simple Faraday rotating medium ($\psi\propto\lambda^2$) outlined at the start of $\S$\ref{sec:far_rot}. In these ``Faraday complex'' scenarios, the polarization angle $\psi$ becomes a non-linear function of $\lambda^2$, making the task of reconstructing the magneto-ionic properties difficult. In general, Faraday-complex media must be modeled in order to gain insight into the physics. Models describing the effect of magnetized turbulence and geometry have been presented in influential works by \citet{Burn1966}, \citet{Tribble1991}, \citet{Sokoloff1998} and~\citet{Brentjens2005}. Below we present selected models that have have been used or suggested in the literature and are implemented in the RM-Tools software. All of these models share the property of having closed-form analytic solutions in the $\lambda^2$ domain; in principle, it is also possible to construct a model in the Faraday depth domain and perform an inverse Fourier transform to obtain the model in the wavelength domain. Table~\ref{tab:models} summarizes the list of models currently included in RM-Tools.

\begin{table*}[]
    \centering
    \begin{tabular}{|c|c|c|c|}
    \hline
     Model \# & Model Name & Equation \# & Model parameters \\ \hline
     m1 & Single Faraday-thin & \ref{eqn:Faraday_model_thin} & $p_0$, $\phi$, $\psi_0$ \\
    m11 & Double Faraday-thin & \ref{eqn:Faraday_model_thin} + \ref{eqn:multiple} & \{$p_0$, $\phi$m $\psi_{0}$\}$_{1,2}$ \\
    m111 & Triple Faraday-thin & \ref{eqn:Faraday_model_thin} + \ref{eqn:multiple} & \{$p_0$, $\phi$, $\psi_{0}$\}$_{1,2,3}$ \\
    m2 & Single turbulent screen & \ref{eqn:Burn_law} & $p_0$, $\phi$, $\psi_0$, $\sigma_\mathrm{RM}$ \\
    m3 & Double turbulent screen, common turbulence & \ref{eqn:Burn_law} + \ref{eqn:multiple} & \{$p_0$, $\phi$, $\psi_{0}$\}$_{1,2}$, $\sigma_\mathrm{RM}$ \\
    m4 & Double turbulent screen, separate turbulence & \ref{eqn:Burn_law} + \ref{eqn:multiple} & \{$p_0$, $\phi$, $\psi_{0}$, $\sigma_\mathrm{RM}$\}$_{1,2}$  \\
    m5 & Single uniform medium & \ref{eqn:slab} & $p_0$, $\phi$, $\psi_0$, $\Delta\phi$ \\
    m6 & Double uniform medium & \ref{eqn:slab} + \ref{eqn:multiple} & \{$p_0$, $\phi$, $\psi_{0}$, $\Delta\phi$\}$_{1,2}$ \\
    m7 & Single turbulent mixed medium & \ref{eqn:internal_depol} & $p_0$, $\phi$, $\psi_0$, $\sigma_\mathrm{RM}$, $\Delta\phi$ \\
    m12 & Single turbulent mixed med.~with turbulent screen & \ref{eqn:internal_depol} + \ref{eqn:Burn_law} & $p_0$, $\phi$, $\psi_0$, $\sigma_\mathrm{RM}$, $\Delta\phi$, $\sigma_\mathrm{RM,screen}$ \\
     \hline
    \end{tabular}
    \caption{List of QU-fitting models currently packaged with RM-Tools. The model numbers are used for selecting the desired model in the code.}
    \label{tab:models}
\end{table*}

\subsection{Faraday-thin source}
For a polarized source at a distance $L$ experiencing only rotation by a foreground screen, the complex polarized signal can be expressed as
\begin{equation}\label{eqn:Faraday_model_thin}
  \tilde{\mathcal{P}}(\lambda^2) = p_0\,e^{2i\,(\psi_0 + \phi\,\lambda^2)},
\end{equation}where $p_0$ is the intrinsic fractional linear polarization, $\psi_0$ is the intrinsic linear polarization angle and $\phi(L)\equiv{\rm RM}$ is the Faraday depth.
This is the simplest line-of-sight model, with a single (simple) source of polarized emission and no (unresolved) spatial structure in the source's polarized emission or the Faraday rotation.

\subsection{Turbulent Faraday screen}

If $\boldsymbol{B}$ or $n_e$ vary on small spatial scales, closely spaced lines of sight will experience different amounts of Faraday rotation. When averaged over a telescope beam, the result is a type of {\it beam depolarization}, called external Faraday dispersion (as it occurs external to the polarization-emitting volume), described by the equation \citep{Burn1966, Tribble1991, Sokoloff1998}
\begin{equation}\label{eqn:Burn_law}
  \tilde{\mathcal{P}}(\lambda^2) = p_0\,e^{2i\,(\psi_0 + \phi\,\lambda^2)}~e^{-2\sigma_{\rm RM}^2\,\lambda^4}.
\end{equation}
Here the second exponential term results in a decrease in polarized fraction towards longer $\lambda^2$ values. The parameter $\sigma_{\rm RM}$ characterizes the Faraday depth fluctuations on scales much smaller than the resolution and hence the degree of depolarization.

\subsection{Uniform mixed emitting and rotating media}
A simple model of a uniform layer of gas containing mixed relativistic and thermal electrons was presented by \citet{Burn1966, Sokoloff1998} and takes the form
\begin{align}\label{eqn:slab}
    & \tilde{\mathcal P}(\lambda^2) =  p_0\,\frac{\rm sin(\Delta \phi\,\lambda^2)}{\Delta \phi\,\lambda^2}\,e^{2i(\psi_{o} + \frac{1}{2}\,\Delta \phi\,\lambda^2)}
\end{align}

Emission from the far side will experience a different amount of rotation compared to emission from the near side, leading to {\it differential Faraday rotation} (also known as {\it depth depolarization} or {\it front-back depolarization}). In Eqn.~\ref{eqn:slab}, $\Delta \phi$ is the difference in Faraday rotation caused by the full path length through the slab (i.e., the Faraday rotation experienced by emission produced at the back side of the region by the point it reaches the front of the region). Such uniform layers have been called `Burn slabs' (after \citealt{Burn1966}) or `uniform slabs' in the literature. When fitting Burn slab models it is common to apply a (uniform) foreground Faraday rotation term, $\exp(2i \phi_{fg} \lambda^2)$, and to make the substitution $\phi_0 = \phi_{fg} + \frac{1}{2} \Delta \phi$, where $\phi_0$ is the mean degree of Faraday rotation experienced by emission from the slab. \citet{Sokoloff1998} proposed a model with multiple layers of uniform slabs (with physical parameters within each layer) which is the summation of the polarization of each layer and with the emission from each more distant layer experiencing the Faraday rotation caused by its foreground layers.

\subsection{Turbulent mixed media}
The uniform slab model described above was also extended by \citet{Burn1966} and \citet{Sokoloff1998} to include turbulent fluctuations in the emitting and Faraday-rotating medium, both along the line of sight and in the plane of sky on scales smaller than the beam. The resulting model has the form
\begin{equation}\label{eqn:internal_depol}
  \tilde{\mathcal{P}}(\lambda^2) = p_0\,e^{2i\,\psi_0}~\frac{1-e^{-\tilde{S}}}{\tilde{S}}
\end{equation}
where $\tilde{S} = 2\sigma_{RM}^2 \lambda^4 - 2i \Delta\phi \;\lambda^2$ contains a real part, which describes the depolarization caused by the turbulent magnetic field, and an imaginary part, which describes the mean Faraday rotation associated with a large-scale magnetic field component. Note that the $\sigma_{RM}$ parameter is generally given the same symbol as the parameter in the external Faraday dispersion model, due to the similarity in the functional forms of the models, but the physical interpretation is different. This model is often referred to as {\it internal Faraday dispersion}, as the turbulent Faraday rotation effect is internal to the emitting volume.

\subsection{Multiple unresolved clumps}\label{sec:clumps}
If $N$ distinct regions of polarization-emitting volumes are present along a line of sight (or are present within a single beam of a particular observation), the observed polarized signal can be constructed by summing the individual models \begin{equation}\label{eqn:multiple} 
\tilde{\mathcal{P}}(\lambda^2) = \tilde{\mathcal{P}}_1 + \tilde{\mathcal{P}_2} + \ldots + \tilde{\mathcal{P}_{N}}.
\end{equation}
Even in the simple case of two Faraday-thin sources, $\tilde{\mathcal{P}}(\lambda^2)$ can exhibit very complex behaviour due to the Faraday components interfering to produce a `beating' pattern. In the limit of low signal-to-noise and narrow bandwidths, discriminating between alternative models of a polarized source, its internal Faraday rotation and foreground Faraday active material can be extremely challenging \citep{Farnsworth2011, Sun2015}.

\section{Performance of different samplers for QU-fitting}\label{sec:sampler_qufit} 

As mentioned in \S\ref{sec:QUfitting}, the QU-fitting script of RM-Tools incorporates the {\it bilby} Python package, which grants flexibility in selecting different Bayesian inference samplers. We assessed the performance of several nested sampling samplers -- {\it dynesty} \citep{Speagle2020}, {\it nestle}\footnote{\href{https://github.com/kbarbary/nestle}{https://github.com/kbarbary/nestle}}, {\it pymultinest} \citep{Buchner2014}, and {\it ultranest} \citep{Buchner2021} -- particularly focusing on the consistency of the output results.

We generated simulated data of two polarized sources observed with the POSSUM Band 1 frequency setup (i.e., 800--1088\,MHz with 1\,MHz channels). The first source (Faraday simple; model m1) was defined as $p_0 = 0.3$, $\phi = -250\,{\rm rad\,m}^{-2}$, and $\psi_0 = +55^\circ$, with a per-channel signal-to-noise of $30$ in polarization. The second source (Faraday complex; model m11) was generated using $p_{0,1} = 0.2$, $\phi_1 = +400\,{\rm rad\,m}^{-2}$, $\psi_{0,1} = +30^\circ$, $p_{0,2} = 0.1$, $\phi_2 = -250\,{\rm rad\,m}^{-2}$, and $\psi_{0,2} = +30^\circ$, with a per-channel signal-to-noise of 10--30 in polarization. In both cases, the total intensity was constant across frequency at a per-channel signal-to-noise of 100.

\begin{figure}
  \centering
  \includegraphics[width=0.32\textwidth, trim=0 0 0 0]{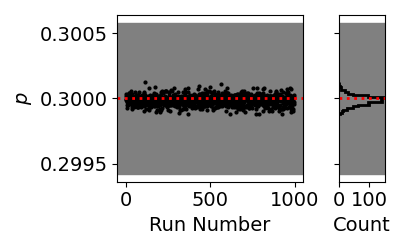}
  \includegraphics[width=0.32\textwidth, trim=0 0 0 0]{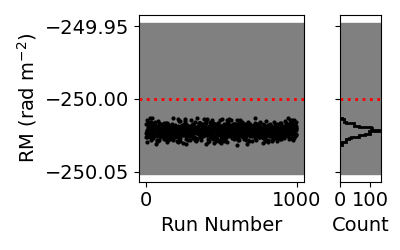}
  \includegraphics[width=0.32\textwidth, trim=0 0 0 0]{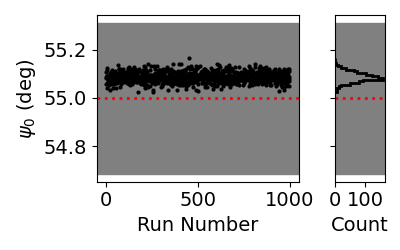}
  \includegraphics[width=0.32\textwidth, trim=0 0 0 0]{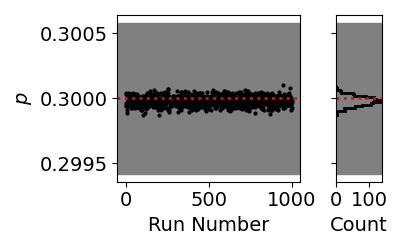}
  \includegraphics[width=0.32\textwidth, trim=0 0 0 0]{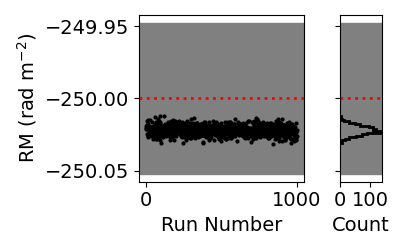}
  \includegraphics[width=0.32\textwidth, trim=0 0 0 0]{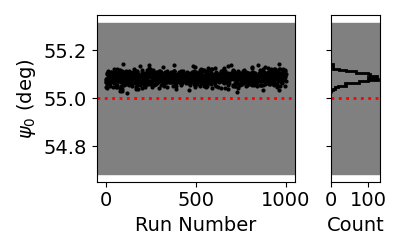}
  \includegraphics[width=0.32\textwidth, trim=0 0 0 0]{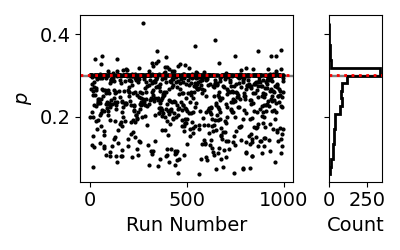}
  \includegraphics[width=0.32\textwidth, trim=0 0 0 0]{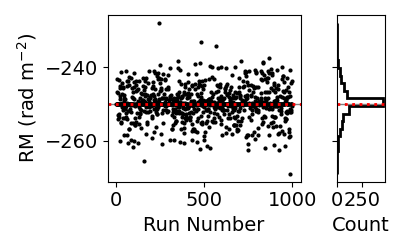}
  \includegraphics[width=0.32\textwidth, trim=0 0 0 0]{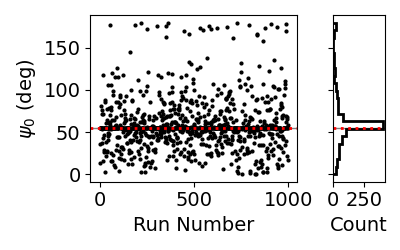}
  \includegraphics[width=0.32\textwidth, trim=0 0 0 0]{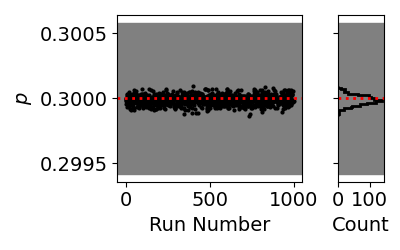}
  \includegraphics[width=0.32\textwidth, trim=0 0 0 0]{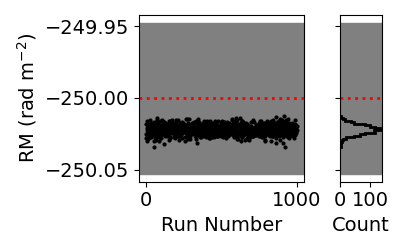}
  \includegraphics[width=0.32\textwidth, trim=0 0 0 0]{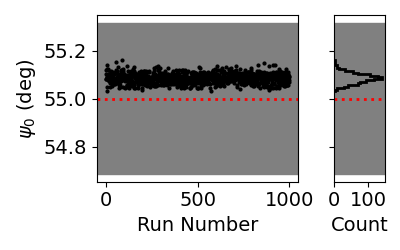}
  \caption{\small Results of performing QU-fitting on the simulated Faraday simple source 1000 times. Each row represents a different nested sampling sampler used (top to bottom): \textit{dynesty}, \textit{nestle}, \textit{pymultinest}, and \textit{ultranest}, and the fitted parameters are divided into the three columns. The red dotted lines mark the parameters' input values, and the grey shaded regions show the median output $1\sigma$ uncertainty.}
  \label{fig:sampler_test_m1}
\end{figure}

\begin{figure}
  \centering
  \includegraphics[width=0.32\textwidth, trim=0 0 0 0]{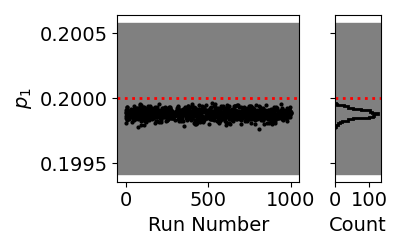}
  \includegraphics[width=0.32\textwidth, trim=0 0 0 0]{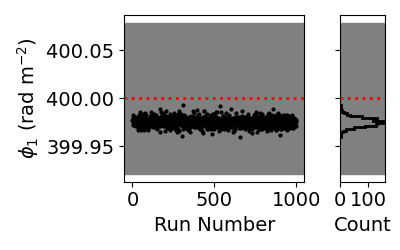}
  \includegraphics[width=0.32\textwidth, trim=0 0 0 0]{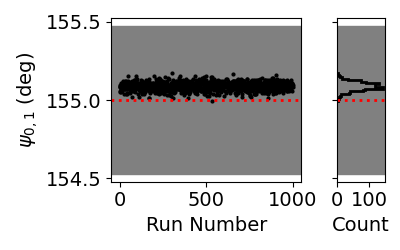}
  \includegraphics[width=0.32\textwidth, trim=0 0 0 0]{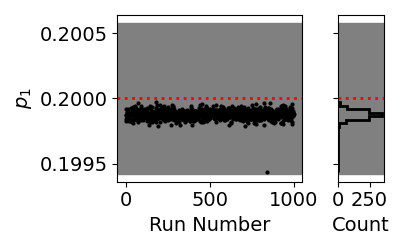}
  \includegraphics[width=0.32\textwidth, trim=0 0 0 0]{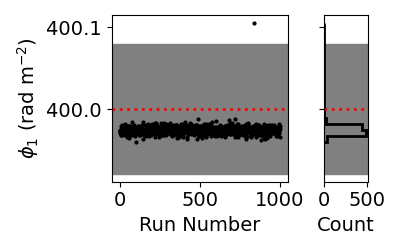}
  \includegraphics[width=0.32\textwidth, trim=0 0 0 0]{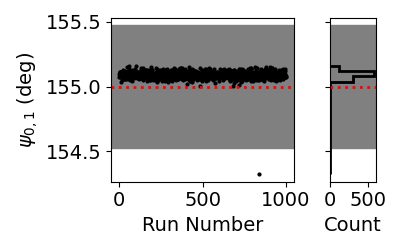}
  \includegraphics[width=0.32\textwidth, trim=0 0 0 0]{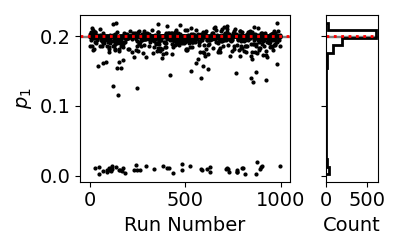}
  \includegraphics[width=0.32\textwidth, trim=0 0 0 0]{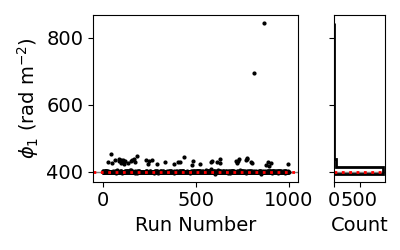}
  \includegraphics[width=0.32\textwidth, trim=0 0 0 0]{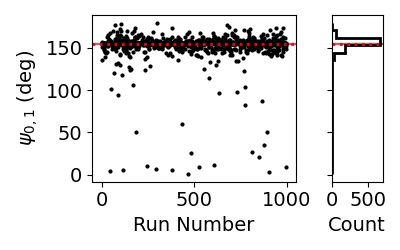}
  \caption{\small Similar to Figure~\ref{fig:sampler_test_m1}, but showing the results from the Faraday complex source. The first Faraday component is shown here, and the second component is shown in Figure~\ref{fig:sampler_test_m11_2}. The \textit{ultranest} sampler is omitted here, since it crashed without converging.}
  \label{fig:sampler_test_m11_1}
\end{figure}

\begin{figure}
  \centering
  \includegraphics[width=0.32\textwidth, trim=0 0 0 0]{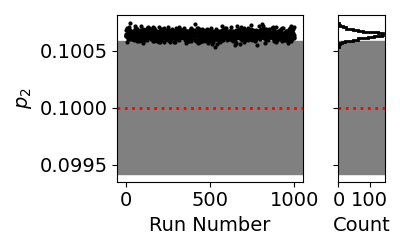}
  \includegraphics[width=0.32\textwidth, trim=0 0 0 0]{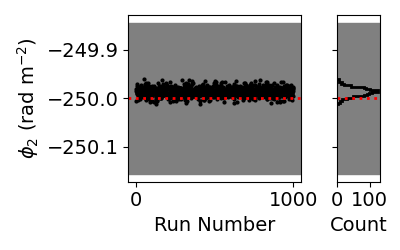}
  \includegraphics[width=0.32\textwidth, trim=0 0 0 0]{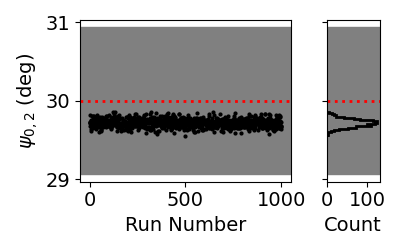}
  \includegraphics[width=0.32\textwidth, trim=0 0 0 0]{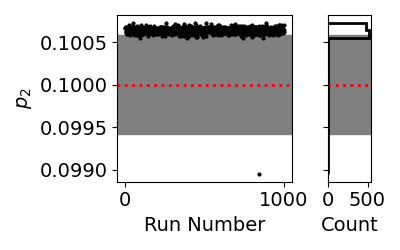}
  \includegraphics[width=0.32\textwidth, trim=0 0 0 0]{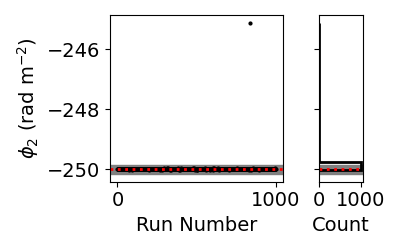}
  \includegraphics[width=0.32\textwidth, trim=0 0 0 0]{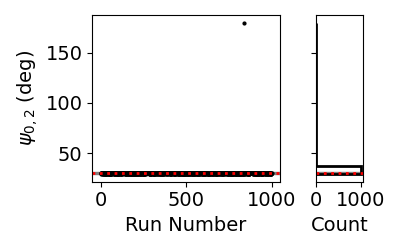}
  \includegraphics[width=0.32\textwidth, trim=0 0 0 0]{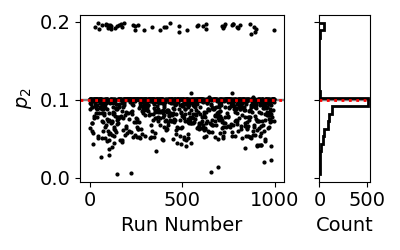}
  \includegraphics[width=0.32\textwidth, trim=0 0 0 0]{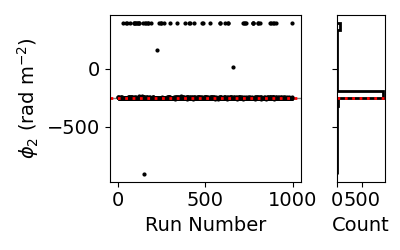}
  \includegraphics[width=0.32\textwidth, trim=0 0 0 0]{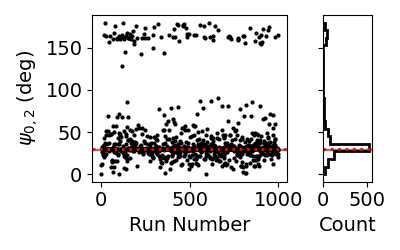}
  \caption{\small As for Figure~\ref{fig:sampler_test_m11_1}, but showing the second Faraday component of the Faraday complex source.}
  \label{fig:sampler_test_m11_2}
\end{figure}

With the two simulated sources above, we performed QU-fitting using the four different nested sampling samplers, repeated 1000 times for each source-sampler combination. The number of active points (\texttt{nlive}) was set to 128 for all cases. Across the 4000 runs per source, the input data were identical (i.e., the injected noise was fixed between runs, instead of being generated for each QU-fitting run). We expect the output values to be similar between runs, but not identical given the stochastic nature of nested sampling. The output results for the Faraday simple source are shown in Figure~\ref{fig:sampler_test_m1}. We found that for \textit{dynesty}, \textit{nestle}, and \textit{ultranest}, the outputs were highly consistent between runs (with differences significantly smaller than the measurement uncertainty shown by the grey shaded area). The results between the three samplers also agree well with each other. We found the outputs from \textit{pymultinest} to vary significantly between runs. The results for the Faraday complex source are shown in Figures~\ref{fig:sampler_test_m11_1}~and~\ref{fig:sampler_test_m11_2}, but with \textit{ultranest} omitted because we did not manage to get even a single complete fitting (the program consistently crashed by overflowing available memory). We found again that \textit{dynesty} and \textit{nestle} both gave highly consistent results between runs, but with a single stray data point from \textit{nestle}. The outputs from \textit{pymultinest} were again highly inconsistent between runs.

We also briefly tested if these were reproduced in the low signal-to-noise regime. We repeated the investigation for both the Faraday simple and Faraday complex cases above, but with a per-channel signal-to-noise in Stokes {\it Q} and {\it U} of 1 for the former, and 0.67--2 for the latter. We found that the results are consistent with the higher signal-to-noise cases above.

Finally, for each of the two sources in the original (high signal-to-noise) case, we further generated two sets of simulated data with different random number generator seeds for the injected noise. The above analysis was then repeated, and we found consistent results as above. In these two follow-up runs, we did not find any stray data points from \textit{nestle} for the Faraday complex source.

With respect to the computational cost of QU-fitting, we do not provide the absolute computational time of our tests, since it can heavily depend on factors such as the sampled parameter ranges, the input data size, the model used, the \texttt{nlive} value used, and the number of compute cores used. Instead, we summarize the approximate relative compute time amongst the samplers, which can be useful for the prospective RM-Tools users to decide on the sampler that they utilize. For the Faraday simple (m1) case, all four samplers ran at a similar rate, with \textit{dynesty} and \textit{pymultinest} being about 50\,\% slower than \textit{nestle} and \textit{ultranest}. Meanwhile for the Faraday complex (m11) case, \textit{pymultinest} was the fastest (albeit with highly inconsistent results, as pointed out above), with \textit{dynesty} being about 10 -- 15 times slower, and \textit{nestle} being about 2--40 times slower than \textit{pymultinest}.

In summary, our tests found that the \textit{pymultinest} sampler can give inconsistent, often sub-optimal, ``best-fit'' results while \textit{ultranest} does not appear to work at all for the m11 model. The \textit{dynesty} and \textit{nestle} samplers work better on the cases we tested, especially the former given its consistent compute efficiency. However, we acknowledge that the performance of the samplers can vary based on factors including (but not limited to) the observation frequency coverage, the parameter values representing the target source, the model used, and the model that can best represent the target source. While we cannot provide an exhaustive test here, we make available a software package\footnote{\href{https://github.com/jackieykma/qufit_sampler_test}{https://github.com/jackieykma/qufit\_sampler\_test}} that facilitates further tests of the performance of the RM-Tools QU-fitting algorithm.




\end{document}